\documentclass[twocolumn,traditabstract,a4]{aa}

\usepackage{times}
\usepackage{graphicx}
\usepackage{subfigure}
\usepackage{amsmath}
\usepackage{color}
\usepackage{multirow}
\usepackage{multicol}
\usepackage{float}
\usepackage{tabularx}
\usepackage{soul}
\usepackage{cancel}
\usepackage{fontawesome}
\bibpunct{(}{)}{;}{a}{}{,} 
\usepackage[colorlinks=true,linkcolor=red,citecolor=blue, breaklinks=true]{hyperref}
\usepackage{natbib,twoopt}
\usepackage{url}
\makeatletter
\newcommandtwoopt{\citeads}[3][][]{\href{http://adsabs.harvard.edu/abs/#3}%
{\def\hyper@linkstart##1##2{}%
\let\hyper@linkend\@empty\citealp[#1][#2]{#3}}}
\newcommandtwoopt{\citepads}[3][][]{\href{http://adsabs.harvard.edu/abs/#3}%
{\def\hyper@linkstart##1##2{}%
\let\hyper@linkend\@empty\citep[#1][#2]{#3}}}
\newcommandtwoopt{\citetads}[3][][]{\href{http://adsabs.harvard.edu/abs/#3}%
{\def\hyper@linkstart##1##2{}%
\let\hyper@linkend\@empty\citet[#1][#2]{#3}}}
\newcommandtwoopt{\citeyearads}[3][][]%
{\href{http://adsabs.harvard.edu/abs/#3}
{\def\hyper@linkstart##1##2{}%
\let\hyper@linkend\@empty\citeyear[#1][#2]{#3}}}
\makeatother

\usepackage{graphicx}
\usepackage{txfonts}
\usepackage{xcolor}
\usepackage{siunitx}
\graphicspath{{Figures/}}

\def\aap{A\&A}

\def\apjs{ApJS}
\def\apjl{ApJL}

\newcommand{\pysm}{\ensuremath{\tt PySM}}
\newcommand{\Nside}{\ensuremath{ N_{\rm side}}}
\newcommand{\planck}{{\textit{Planck}\/}}
\newcommand{\wmap}{{ WMAP\/}}
\newcommand{\quijote}{{ QUIJOTE\/}}
\newcommand{\healpix}{\ensuremath{\tt HEALPix}}

\newcommand{\commander}{\ensuremath{\tt Commander}}
\newcommand{\cosmoglobe}{\ensuremath{\tt COSMOGLOBE}}

\def\GHz{\ifmmode $\,GHz$\else \,GHz\fi}
\def\MHz{\ifmmode $\,MHz$\else \,MHz\fi}
\def\mKs{\ifmmode $\,mK\,s$^{1/2}\else \,mK\,s$^{1/2}$\fi}
\def\muKs{\ifmmode \,\mu$K\,s$^{1/2}\else \,$\mu$K\,s$^{1/2}$\fi}

\begin{document}

\title{QUIJOTE scientific results}
\subtitle{XIX. New constraints on the synchrotron spectral index using a semi-blind component separation method}
\titlerunning{Synchrotron spectral index with semi-blind method}

\author
{ Debabrata Adak,\inst{1,2}\thanks{E-mail: adak@iac.es}
\and
J.~A.~Rubi\~{n}o-Mart\'{\i}n\inst{1,2}
\and
R.~T.~G\'{e}nova-Santos\inst{1,2}
\and
M.~Remazeilles\inst{3}
\and
A. Almeida\inst{1,2}
\and
K. Aryan\inst{1,2}
\and
M.~Ashdown\inst{4,5}
\and
R.~B.~Barreiro\inst{3}
\and
U. Bose\inst{1,2}
\and
R. Cepeda-Arroita\inst{1,2}
\and
J.M. Casas\inst{1,2}
\and
M.~Fern\'{a}ndez-Torreiro\inst{6}
\and
E.~Mart\'{i}nez-Gonzalez\inst{3}
\and
F.~Poidevin\inst{1,2}
\and
R.~Rebolo\inst{1,2,7}
\and
P.~Vielva\inst{3}
}

\institute{Instituto de Astrof\'{\i}sica de Canarias, E-38200 La Laguna, Tenerife, Spain
\and
Departamento de Astrof\'{\i}sica, Universidad de La Laguna, E-38206 La Laguna, Tenerife, Spain
\and
Instituto de F\'{\i}sica de Cantabria (IFCA), CSIC-Univ. de Cantabria, Avda. los
Castros, s/n, E-39005 Santander, Spain
\and
Astrophysics Group, Cavendish Laboratory, University of Cambridge, J J Thomson Avenue, Cambridge CB3 0HE, UK
\and
Kavli Institute for Cosmology, University of Cambridge, Madingley Road, Cambridge CB3 0HA, UK
\and
Laboratoire de Physique Subatomique et de Cosmologie, Universit\'{e} Grenoble Alpes, CNRS/IN2P3, 53 Avenue des Martyrs, Grenoble, France
\and
Consejo Superior de Investigaciones Cient\'{\i}ficas, E-28006 Madrid, Spain
}

\date{Received 20 10 2025; accepted 02 02 2026}

\abstract{We introduce a novel approach to estimate the spectral index, $\beta_s$, of polarised synchrotron emission,  combining the moment expansion of Cosmic Microwave Background foregrounds and the constrained Internal Linear Combination method. We reconstructed the maps of the first two synchrotron moments, combining multi-frequency data, and applied the `T-T plot' technique between two moment maps to estimate the synchrotron spectral index.  This approach offers a new technique for mapping the foreground spectral parameters, complementing the model-based parametric component separation methods. Applying this technique, we derived a new constraint on the spectral index of polarised synchrotron emission using QUIJOTE  MFI wide-survey 11 and 13\,GHz data,  Wilkinson Microwave Anisotropy Probe  data at K and Ka bands, and \planck\ LFI 30\,GHz\ data. 
In the Galactic plane and North Polar Spur regions, we obtained an inverse-variance-weighted mean synchrotron index of $\beta_s = -3.11$ with a standard deviation of $0.21$ due to intrinsic scatter, consistent with  previous results based on parametric methods  using the same dataset. We find that the inverse-variance-weighted mean spectral index, including both statistical and systematic uncertainties, is $\beta_s^{\rm plane} = -3.05 \pm 0.01$ in the Galactic plane and $\beta_s^{\rm high\text{-}lat} = -3.13 \pm 0.02$ at high latitudes, indicating a moderate steepening of the spectral index from low to high Galactic latitudes. Our analysis indicates that, within the current upper limit on the Anomalous Microwave Emission  polarisation fraction, our results are not subject to any appreciable bias.
Furthermore, we infer the spectral index over the entire QUIJOTE survey region, partitioning the sky into 21 patches. 
This technique can be further extended to constrain the synchrotron spectral curvature by reconstructing higher-order moments when better-quality data become available.
}

\keywords{methods: cosmology: observations -- cosmic microwave background -- Polarised foregrounds --- synchrotron emission}
\authorrunning{D. Adak et al.}
\maketitle

\section{Introduction}\label{sec:intro}
The increasing sensitivity of Cosmic Microwave Background (CMB) observations has revolutionised our understanding of the Universe in the past three decades. With the tremendous success of three space-based CMB missions, Cosmic Background Explorer \citep{COBE:1990}, Wilkinson Microwave Anisotropy Probe (WMAP) \citep{Bennett_2013}, and \planck\ \citep{Planck-I:2020}, and plenty of ground-based and sub-orbital CMB experiments \citep{SPT:2011PASP,2024arXiv240519469K,ACT:2024A,SPIDER:Ade_2025}, we have reached an era where the primary focus of the cosmological observations is to understand the polarisation of the CMB in the upcoming decades. In particular,  a hunt for the inflationary gravitational waves through the potential detection of the primordial $B$-mode polarisation of CMB is one of the key targets of future CMB polarisation missions \citep{B-mode:review-2016}. Future experiments, including \textit{LiteBIRD} \citep{LiteBIRD:2023}, Simons Observatory \citep{so:2019},  CMB-Bharat \citep{CMB-Bharat:2022}, and PICO \citep{PICO:2019} will provide the most sensitive polarisation maps that will significantly tighten the constraints on the physics of inflation.  

The weak signal of $B$-mode CMB polarisation as compared to the polarised foregrounds makes the experiments very challenging. At frequencies below $\sim$70 \GHz, the dominant foreground contamination is diffuse synchrotron emission generated from gyrating cosmic ray electrons around Galactic magnetic fields \citep{Rybicki-Lightman:1979, Planck-IV:2020}. At frequencies >100 \GHz, polarised thermal emission from dust grains is the key obstacle for $B$-mode observations \citep{Planck-IV:2020}. Therefore, understanding the spectral and spatial behaviour of these diffuse emissions and subtracting them from the data is an obvious step to reliably map the fluctuation of CMB polarisation. Another potential foreground contamination for $B$-mode studies is Anomalous Microwave Emission (AME) seen at the frequency range $\sim$10–60 \GHz\ \citep{Remazeilles:2016, Dickinson:2018rle, Cepeda-Arroita:2025}. The favoured model of AME emission predicts that the radiation is due to electric dipole emission from spinning dust grains of size in the nanometer range \citep{Drain_hensley:2013,Ali:2009} populated at different environments  of the Interstellar medium  \citep{Hoang_alex:2016,Hensley:2022}. However, the true origin of AME emission is still ambiguous. Although AME is a well-motivated foreground component for intensity data, the polarisation property of AME is not well understood from available data. The current constraints on AME polarisation are set to below $1\%$ from various observations \citep{Génova-Santos:2017,Herman:2023,M31quijotemfi,Gonz:2024}. Therefore, it is generally believed that AME is not a major obstacle for CMB $B$-modes. However, it cannot be totally disregarded because: 1) there is still uncertainty on the theoretical models, and 2) the constraints on the polarisation fraction are obtained on specific regions and could be affected by observational effects such as beam depolarisation. The recent studies of \cite{Herman:2023} has found inference of  the AME polarisation fraction depends on the choice of the prior on the synchrotron spectral index, concluding that the current data sets are not strong enough to simultaneously and robustly constrain both polarised synchrotron emission and AME. In this work, therefore, we investigate using simulations whether AME polarisation biases the synchrotron spectral index measurement using our methodology. 

The CMB and various astrophysical foregrounds arise from distinct physical processes and therefore exhibit different spectral behaviours. These spectral differences allow us reliable modelling of the sky emission and enables the extraction of CMB polarisation maps from data obtained by multiple experiments \citep{Eriksen:2008,FG_buster:2009,de_la_Hoz:2023,Brilenkov:2023}. The method requires fitting at least six parameters for the case of the simplest polarised sky model. With the sensitivity of current data, the best-fit results at a significant fraction of the sky are degenerate, prior-dependent, and the signal-to-noise ratio (S/N) is low for various foreground parameters \citep{planck-xxv:2016, C-BASS:2019,de_la_Hoz:2023}. For instance, to reduce the inter-dependency among the foreground spectral parameters, in the main analysis of \planck\ official data using \commander, the synchrotron spectral index has been fixed to the best-ﬁt value derived from the \planck\ 2015 temperature data, corresponding to $\beta_s$ = --3.1 \citep{Planck-IV:2020}. Many attempts have been made to exploit the current available data that yield the synchrotron spectral indices ranging between $-5$ to $-1$ (see \cite{Belsunce:2022} and references therein), which is inconsistent with predictions from the energy distribution of cosmic ray electrons \citep{Rybicki-Lightman:1979,Yang:2017, Neronov:2017}. 
Therefore, exploiting all pieces of information from independent foreground spectrum analysis is always helpful for parametric component separation to set the priors. For example, synchrotron spectral index estimation from experiments such as S-PASS, C-BASS, QUIJOTE, and CLASS \citep{Krachmalnicoff:2018,Dickinson:2019, de_la_Hoz:2023, Watts:2024, Eimer:2024} is useful since their channels are synchrotron-dominated. Furthermore, for the recent development of semi-blind component separation methods \citep{Remazeilles:2021MNRAS.503.2478R, Adak:2021, Carones:2023, Carones:2024},  prior knowledge of the spatial distribution of the spectral index is a piece of very useful information to deproject a few moments \citep{Chluba:2017,Vacher:2023} of the synchrotron to reduce residual leakage to recovered CMB maps.

In this work, we establish a new semi-blind approach to map the synchrotron spectral index distribution using low-frequency data from  WMAP K and Ka bands \citep{Bennett_2013}, \planck\ LFI 30 \GHz\ data \citep{Planck-II:2020}, and QUIJOTE MFI \citep{Rubiño-Martín:2023}, where the synchrotron is the dominant diffuse emission.  For this endeavour, we use a generalised method of constrained Internal Linear Combination \citep[cILC,][]{cILC:2011} for spin-2 fields \citep{PILC:2016,Adak:2021}. The cILC is designed to estimate the map of the desired component from a combination of multi-frequency data, deprojecting other emissions.  In this work, we aim to reconstruct the synchrotron moment maps by applying this technique and estimate the average spectral index from the correlation of the first two moments. As illustrated in Figure~\ref{fig:moments_amp}, the first two synchrotron moments exhibit distinct spectral energy distributions. Consequently, frequency channels below 40 \GHz, where thermal dust emission remains subdominant compared to the synchrotron moments, are expected to aid in disentangling the two moments with this method. In literature, the cILC method has been extensively employed on spin-0 fields  for reconstruction of various cosmological signals \citep{cILC:2011,2020MNRAS.494.5734R, Remazeilles:2021MNRAS.503.2478R,  Rotti:2021, Carones:2023,Carones:2024}. The straightforward generalisation of this method for spin-2 fields is to decompose the $Q, U$ maps to $E$ and $B$-mode polarisation and apply cILC on $E$ and $B$-mode maps separately. However, the decomposition of the $Q, U$ maps to $E$, $B$-mode polarisation maps on incomplete sky data leads to leakage of $E$ to $B$-mode. This motivates us to apply the constrained polarisation ILC (cPILC hereafter, \cite{Adak:2023}) directly on $Q, U$ maps instead.  The cPILC method was developed in \cite{Adak:2023}, which is a generalisation of the polarisation ILC (PILC) method developed in \cite{PILC:2016} for the case of more than one constraint \citep{Remazeilles:2011}. As we are working with Stokes $Q$ and $U$ parameters in Galactic coordinates, one major caveat of the cPILC method is that since it simultaneously minimises $Q$ and $U$ variance, the method is sub-optimal for $U$ mode if the variance is largely driven by $Q$ in general. However, in the case of synchrotron moment recovery, since $U$ has a significant contribution in the variance, the reconstruction is expected to be reasonably good. Furthermore, cPILC is designed to preserve the coherence between the two spinorial components by estimating weights to minimise the variance of a spin-0 field composed of a combination of $Q$ and $U$ \citep{PILC:2016}. On the contrary, recovery of synchrotron moments independently in $Q$ and $U$ requires multiplying the Stokes parameters by different weights. Since  $Q$ and $U$ depend on the local coordinate frame, this implies that weights are determined by changing arbitrarily the polarisation angle and modulus, spoiling the physical meaning of the spin-2 fields. Therefore, in this work, we employ the cPILC method for our purpose. 

This paper is organised as follows. We explain the methodology to estimate the synchrotron moments and inference of spectral index using their correlation in Section~\ref{sec:methodlogy}. In Section~\ref{sec:sims}, we describe the sky models of the simulation used in the validation of the proposed method. In Section~\ref{sec:data}, we describe the data sets used in this work. We validate the method on simulated sky maps and present the inferred results in Section~\ref{sec:sim_val}. In Section~\ref{sec:results}, we provide the inferred spectral index map at \Nside=32 at low and intermediate latitudes where the S/N of synchrotron in data is high and compare our results with the previous analyses in literature. We repeat a similar analysis over the patches defined in \cite{Fuskeland:2014} that include regions of low S/N located at high latitudes and discuss the obtained results in Section~\ref{sec:reg_analysis}. We conclude the results in Section~\ref{sec:summery and conclusion}.

\section{Methodology}\label{sec:methodlogy}
\subsection{Estimation of synchrotron moment maps}\label{sec:moment_est}
The Galactic foreground emissions are considered to be the superposition of emissions from many emission blocks along the line-of-sight (LOS). For the synchrotron and thermal dust, the emission follows a power-law and modified blackbody spectrum (MBB), respectively, with variable spectral parameters along the LOS. Therefore, the integrated total emission along the LOS and within the beam size alters the spectral properties of foregrounds, e.g., the sum of multiple power laws is not a power law, and likewise, the sum of multiple MBB spectra is not an MBB. Considering there is an infinite number of emission blocks along  LOS, foreground emissions are statistically expressed as a summation of infinite moments following the moment expansion method of \cite{Chluba:2017}.
 The synchrotron Stokes parameters $Q^{sync}_{\nu}(p),U^{sync}_{\nu}(p) $ at frequency $\nu$ can be modelled with an amplitude $Q_{\nu_{s}}(p),U_{\nu_{s}}(p)$ at reference frequency $\nu_{s}$ and a statistical average of SEDs along LOS and inside the beam, $f_{sync}(\nu,\beta_s(p))$ = $\int (\nu/\nu_{s})^{\beta_s(p(r))}\mathcal{P}(\beta_s(p(r)))dr$,
\begin{equation}
\label{eq:sync_spec_law}
    \left[\substack{Q^{sync}_{\nu}(p)\\U^{sync}_{\nu}(p)}\right]  = \left[\substack{Q_{\nu_{s}}(p)\\U_{\nu_{s}}(p)}\right]f_{sync}(\nu,\beta_s(p)),
\end{equation}
where $\mathcal{P} $ denotes the probability density of having the spectral index $\beta_s(p(r))$ at pixel $p$ along the LOS distance $r$. Following the moment expansion of synchrotron emission in \cite{Chluba:2017}, we can express Equation~\ref{eq:sync_spec_law} as a summation of moments around some pivot value of the spectral index, $\Bar{\beta}_{s}$, as

\begin{align}
\label{eq:sync_moments}
\left[\substack{Q^{sync}_{ \nu}(p)\\ U^{sync}_{ \nu}(p)}\right] 
&= \left[\substack{Q_{\nu_{s}}(p)\\U_{\nu_{s}}(p)}\right]f_{\rm sync}\left(\nu,\overline{\beta}_s\right)\\ \nonumber
&+ \left[\substack{Q_{\nu_{s}}(p)\\U_{\nu_{s}}(p)}\right]\Delta \beta_s(p)\partial_{{\beta}_s} f_{\rm sync}\left(\nu,\overline{\beta}_s\right) + \mathcal{O}(\beta_{s}^2),             
 \end{align}
where  $\Delta \beta_s(p)=\beta_s(p) - \overline{\beta}_s$,  and 
\begin{subequations}
\begin{flalign}
\label{eq:sync_moments1}
&f_{\rm sync} \left(\nu,\overline{\beta}_s\right) = \left({\nu\over\nu_s}\right)^{\overline{\beta}_s},\\
&\partial_{\overline{\beta}_s} f_{\rm sync} \left(\nu,\overline{\beta}_s\right) = \ln\left({\nu\over\nu_s}\right)f_{\rm sync} \left(\nu,\overline{\beta}_s\right).
\end{flalign}
\end{subequations}
Here we do not consider the higher order moments assuming their contribution can be significantly reduced with a proper choice of pivot value of $\overline{\beta}_s $
\citep{Rotti:2021,Vacher:2023}. 
In principle,  since spatial distribution of the synchrotron spectral index is non-Gaussian, achieving high accuracy in synchrotron modelling using moment expansion requires adding more moments, especially where the spectral index deviates further from its pivot value. However, this also requires deprojection of higher-order moments to reduce bias in the moments we aim to recover. Given the sensitivities of the data used here, this would introduce a noise penalty. Therefore, we restrict our moment expansion up to the first order.

Similarly, we can model the thermal dust emission with moments expressed with respect to the pivot values of dust temperature, $\overline{T}_d$, and spectral index, $\overline{\beta}_d$, as 
\begin{align}
\label{eq:dust_moments}
[Q^{\rm dust}_{\nu}(p), U^{\rm dust}_{\nu}(p)]&=[Q_{\nu_d}(p), U_{\nu_d}(p)]f_{dust}(\nu,T_d (p),\beta_d (p))\cr 
&=[Q_{\nu_d}(p), U_{\nu_d}(p)]f_{\rm dust}\left(\nu,\overline{\beta}_d, \overline{T}_{\!d}\right)  \nonumber \cr 
&+ [Q_{\nu_d}(p), U_{\nu_d}(p)]\Delta \beta_d(p)\partial_{{\beta}_d} f_{\rm dust}\left(\nu,\overline{\beta}_d, \overline{T}_{\!d}\right) \\ \nonumber
&+ [Q_{\nu_d}(p), U_{\nu_d}(p)]\Delta T_d(p)\partial_{{T}_d} f_{\rm dust} \left(\nu,\overline{\beta}_d, \overline{T}_{\!d}\right)\\
&+ .... , 
\end{align}
where $\Delta \beta_{d}(p) = \beta_d (p) - \overline{\beta}_d$, $\Delta T_{d}(p) = T_d (p) - \overline{T}_d$, and 
\begin{flalign}
\label{eq:dust_moments1}
&f_{\rm dust} \left(\nu,\overline{\beta}_d, \overline{T}_{\!d}\right) = \left({\nu \over \nu_d}\right)^{\overline{\beta}_d+1} {e^{{\overline{x}_d}}-1\over e^{\overline{x}}-1},\nonumber\\
&\partial_{{\beta}_d} f_{\rm dust} \left(\nu,\overline{\beta}_d, \overline{T}_{\!d}\right) = \ln\left({\nu\over\nu_d}\right)f_{\rm dust} \left(\nu,\overline{\beta}_d, \overline{T}_{\!d}\right),\nonumber\\
&\partial_{{T}_d} f_{\rm dust} \left(\nu,\overline{\beta}_d, \overline{T}_{\!d}\right) = {1\over\overline{T}_{\!d}}\left[{ \overline{x}e^ {\overline{x}}  \over e^ {\overline{x}}  - 1} - { \overline{x}_d e^{\overline{x}_d} \over e^{\overline{x}_d}  - 1 }\right] f_{\rm dust}\left(\nu, \overline{\beta}_d, \overline{T}_{\!d}\right),\nonumber\\
\end{flalign}
are the spectra of different moments. Here, $x_d = {h \nu_d\over K_B T_d}$ and $\overline{x} = {h \nu\over K_B \overline{T}_d}$.

\begin{figure}
    \centering
    \includegraphics[width=\linewidth]{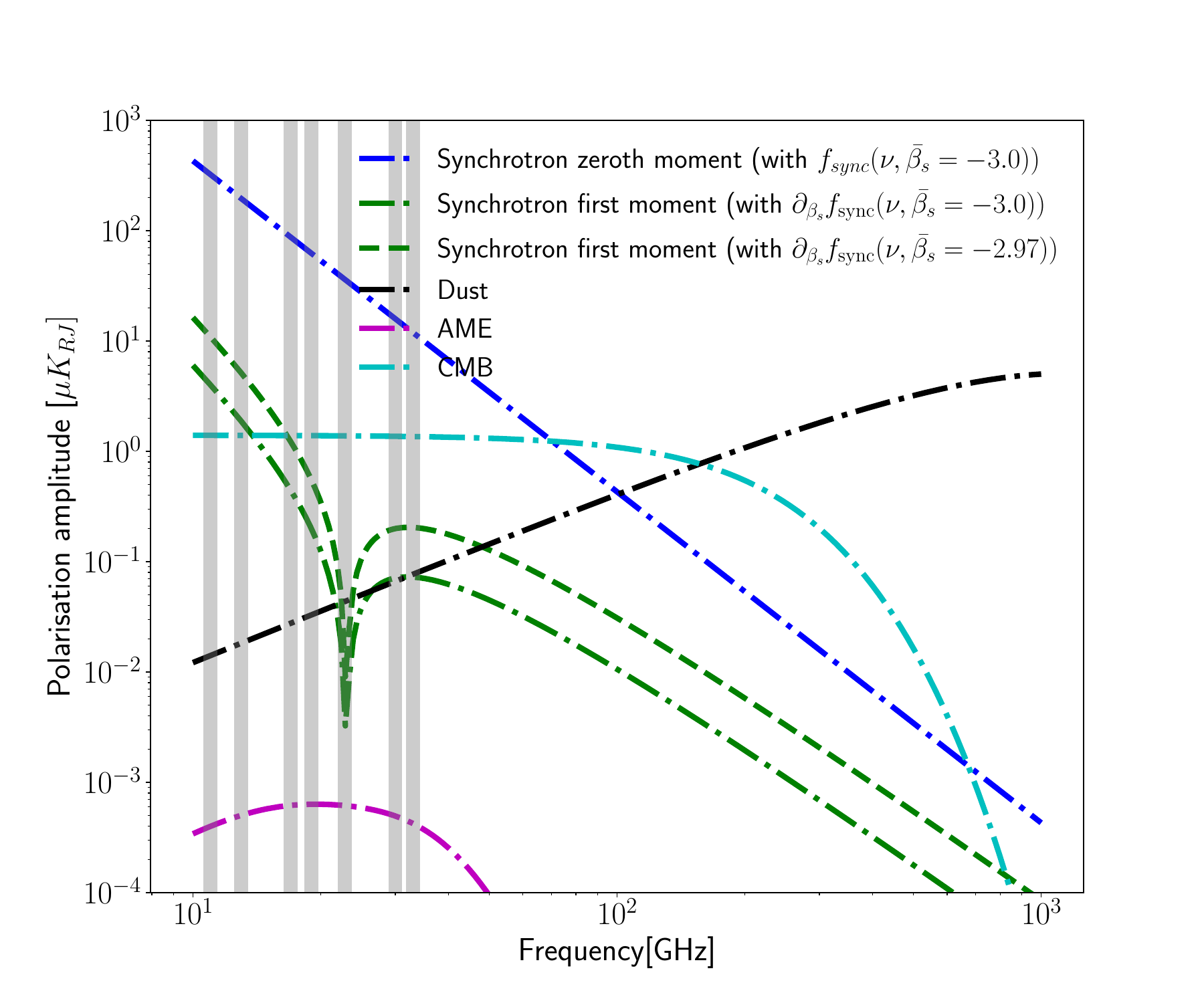}
    \caption{Polarisation intensity, $P = \sqrt{Q^{2} + U^{2}}$, of the synchrotron zeroth moment (dash-dotted blue line) and the first moment (green) for two pivot values of $\beta_s$ = --3.0 (dash-dotted green line) and --2.97 (dashed green lines) at a pivot frequency, $\nu_s = 22.8$ \GHz. Emissions from dust, CMB, and AME (0.5 \% polarisation fraction) are presented by black, cyan, and magenta dash-dotted lines, respectively. The grey areas represent the frequency bands of QUIJOTE MFI bands, WMAP K and Ka-band, and \planck\ LFI 30 GHz. }
    \label{fig:moments_amp}
\end{figure}

The spin-2 fields, $P^{\pm} (p) = Q(p) \pm iU(p)$ of the microwave polarisation data at frequency $\nu$ can be described as
\begin{align}
\label{eq:data_representation}  
    P^{\pm}_{\nu}(p) &= P^{\pm}_{\nu_{s}}(p)f_{\rm sync} \left(\nu,\overline{\beta}_s\right) + P^{\pm}_{\nu_{s}}(p)\Delta \beta_s(p) \partial_{\overline{\beta}_s} f_{\rm sync} \left(\nu,\overline{\beta}_s\right) + n^{\pm}_{\nu}(p)\nonumber\\ 
    &=\sum_{ c=1}^{N_c} A_{\nu c} P^{\pm}_c (p) + n^{\pm}_{\nu} (p),
\end{align}
where $A_{\nu c}$ is the mixing matrix associated with the components $P^{\pm}_{c}(p)$, $N_{c} $ is the number of components and  $n^{\pm}_{\nu} (p)$ comprises the total nuisance emission from CMB, dust moments, AME and instrument noise. In order to take into account the effect of instrument bandpass in real data, we perform a bandpass integration over the respective spectrum of the components while computing the mixing matrix in real data analysis. This prescription is complementary of multiplying colour correction factors to the data. In Equation~\ref{eq:data_representation}, we can consider the spin-2 fields of synchrotron amplitude or zeroth moment, $P^{\pm}_{\nu_{s}}(p)$, and first moment, $P^{\pm}_{\nu_{s}}(p)\Delta \beta_{s}(p)$ to be two independent components following the spectrum, $f_{\rm sync} \left(\nu,\overline{\beta}_s\right)$ and $\partial_{\overline{\beta}_s} f_{\rm sync} \left(\nu,\overline{\beta}_s\right)$ respectively. 
In Figure~\ref{fig:moments_amp}, we display polarisation intensity ${P} = \sqrt{Q^{2} + U^{2}}$ of synchrotron zeroth moment (dash-dotted blue line) and first moments for two choice of pivot spectral parameter, $\overline{\beta}_s$ = -- 3.0 (dash-dotted green line) and --2.97 (dashed green line) and pivot frequency $\nu_s = 22.8$ \GHz. We also present other nuisance emission from CMB (cyan), thermal dust (black) and AME (magenta) with 0.5\% polarisation fraction. 

In this work, our first step is to estimate the \textit{zeroth moment} or synchrotron amplitude, $P^{\pm}_{\nu_s}(p)$ and the \textit{first moment} which is $P^{\pm}_{\nu_s}(p)\Delta\beta_s(p)$.
Since the first two moments of synchrotron emission are highly correlated but follows two different spectrum as shown in Figure~\ref{fig:moments_amp}, in order to deproject the possible contamination from one component while estimating another as well as other nuisance foregrounds, we apply the cPILC  method as motivated in Section~\ref{sec:intro}.  Here we briefly summarise the method. 

Assuming the mixing matrix to be constant over some domain $\mathcal{D}(p)$\footnote{Here we consider the whole region shown in Figure~\ref{fig:high_snr_mask} as a single domain since we are not changing the pivot values of the dust and synchrotron spectral parameters.}  of the sky for a reasonable choice of the pivot values of the synchrotron and dust spectral parameters, Equation~\ref{eq:data_representation} for the data vector $\textbf{P}^{\pm} (p) = \{P^{\pm}_{\nu}\}(p)$ at all channels can be generalised as
\begin{equation}  
\label{eq:data_vec}
    \textbf{P}^{\pm}  (p) = \textbf{A} \boldsymbol{P}_c^{\pm}  (p) + \textbf{n}^{\pm} (p). 
\end{equation}
The estimation of the desired component $P^{\pm}_c$ deprojecting others in the cPILC method consists of a weighted sum of available channel maps,
\begin{equation}
\label{eq:eqtimated_comp}
    \hat{P}_c^{\pm} (p) = \boldsymbol{w}^T\textbf{P}^{\pm}(p).
\end{equation}
The weights $\textbf{\textit{w}} = \{w_{\nu}\}$ are determined by minimising  the covariance matrix, $C_{\nu \nu^{'}} = \left\langle Q_{\nu} (p) Q_{\nu^{'}} (p) +  U_{\nu} (p) U_{\nu^{'}} (p) \right\rangle$ estimated using the all pixels within domain  $\mathcal{D}(p)$ while the weights have unit response to the spectrum of the component of interest and zero response to the spectrum of the components desired to deproject. In particular, for estimation of first synchrotron moment in Equation~\ref{eq:data_representation}, the conditions applied to the weights are:
\begin{subequations}
\begin{equation}
\label{eq:constraints_on_weights_a}
    \textbf{\textit{w}}^T\partial_{\overline{\beta}_s} \textbf{\textit{f}}_{\rm sync}  = 1 ,
\end{equation}   
\begin{equation} 
\label{eq:constraints_on_weights_b}
     \textbf{\textit{w}}^T \textbf{\textit{f}}_{\rm sync}  = 0,
\end{equation}
\begin{equation} 
\label{eq:constraints_on_weights_c}
     \textbf{\textit{w}}^T \textbf{\textit{f}}_{\rm dust}  = 0.
\end{equation}
\end{subequations}
In order to minimise possible contamination from thermal dust, we deproject its zeroth moment by applying the constraint given in Equation~\ref{eq:constraints_on_weights_c}. Introducing additional constraints can further reduce contamination from higher-order moments, but this comes at the cost of increased noise in the recovered map \citep{Remazeilles:2021MNRAS.503.2478R}. Hence, the number of imposed constraints must be chosen carefully to balance the trade-off between residual foreground contamination and  residual noise amplification. In our specific case, the higher-order moments of the synchrotron component are several orders of magnitude smaller than the first two moments, and thermal dust contributes negligibly within the frequency range considered. Therefore, we do not impose additional constraints to deproject higher-order moments of either component.
The condition of minimum covariance of $C_{\nu \nu^{'}} $  with a set of constraints in Equations~\ref{eq:constraints_on_weights_a}, \ref{eq:constraints_on_weights_b}, and \ref{eq:constraints_on_weights_c} using  the Lagrange's undetermined multipliers method yields the optimised weights,
\begin{equation}
\label{eq:weights_firts_moments}
\textbf{\textit{w}}^{T} = \textbf{e}^T(\textbf{F}^T\textbf{C}^{-1}\textbf{F})^{-1}\textbf{F}^{T}\textbf{C}^{-1},
\end{equation}  
where $\textbf{F} =[\partial_{\overline{\beta}_s} \textbf{\textit{f}}_{\rm sync},\,  \textbf{\textit{f}}_{\rm sync}, \,\textbf{\textit{f}}_{\rm dust}]$ and $\textbf{e} =[1,0,0]^{T}$. The minimum variance estimated first synchrotron moment is then given by
\begin{equation}
\label{eq:estimated_firts_moment}
    \hat{s}^{\pm}_1(p) = P^{\pm}_{\nu_{s}} (p)\Delta \beta_{s}(p) + \textbf{\textit{w}}^{T}\textbf{n}^{\pm}(p), 
\end{equation}
which is an unbiased estimation of the first synchrotron moment with zero contamination of synchrotron amplitude or zeroth moment and some residual contamination from remaining nuisance components in Equation~\ref{eq:data_representation}. Similarly, if we interchange the constraints in Equations~\ref{eq:constraints_on_weights_a}  and ~\ref{eq:constraints_on_weights_b}, we can obtain the unbiased estimation of the zeroth synchrotron moment, 
\begin{equation}
\label{eq:estimated_zeroth_moment}
    \hat{s}^{\pm}_0(p) = P^{\pm}_{\nu_{s}} (p) + \overline{\textbf{\textit{{w}}}}^T\textbf{n}^{\pm}(p),
\end{equation}
From Equations~\ref{eq:estimated_firts_moment} and ~\ref{eq:estimated_zeroth_moment}, we finally estimate the map of spectral index modulated synchrotron amplitude, $P^{\pm}_{\nu_s} \beta_{s} (p)$ following

\begin{equation}
\label{eq:combo_zeroth_firt_moment}
    \hat{s}^{\pm} (p) = \hat{s}_{1}^{\pm} (p) + \overline{\beta}_s\hat{s}_{0}^{\pm} (p) = P^{\pm}_{\nu_s} \beta_{s} (p) + \hat{\textbf{w}}^T\textbf{n}^{\pm}, 
\end{equation}
where $\hat{\textbf{w}} = \textbf{\textit{w}} + \overline{\beta}_{s}\overline{\textbf{\textit{w}}}$.
We derive the Stokes parameter maps, $Q(p)$ and $U(p)$, from their corresponding recovered spin-2 fields, $\hat{s}_{0}^{\pm}(p)$ and $\hat{s}^{\pm}(p)$, obtained from Equations~\ref{eq:estimated_zeroth_moment} and \ref{eq:combo_zeroth_firt_moment}, respectively. For convenience, we denote these maps as $m_0(p) = \{Q(p), U(p)\}$ and $m_1(p) = \{Q(p)\beta_s(p), U(p)\beta_s(p)\}$, and we will use these notations in the following sections.

\subsection{Estimation of spectral index map}\label{sec:TT_correlation}
The derived Stokes parameters corresponding to two moment maps in Equation~\ref{eq:estimated_zeroth_moment} and ~\ref{eq:combo_zeroth_firt_moment} are related as
\begin{equation}\label{eq:linear_connection}
    m_1(p) + n_1(p) = \hat{\beta}_s m_0(p) + n_0(p),
\end{equation}
where the corresponding noise vector are denoted by $n_0(p)$ and $n_1(p)$ respectively. In Equation~\ref{eq:linear_connection}, we assume both $Q$ and $U$ follow the same average spectral index, $ \hat{\beta}_s$ over some region. We perform the linear regression between two moments using the {`T-T plot'} technique \citep{Turtle:1962} over the sub-domain $\mathcal{S}(p)$ $\in$ $\mathcal{D}(p)$ to estimate the average spectral index within $\mathcal{S}(p)$\footnote{Here we define $\mathcal{S}$ to be the array of all sub-pixels at \Nside\ = 64 inside the  super-pixels at \Nside\ = 32. Therefore, our estimated spectral index map is at \Nside = 32.}. It is worth noticing in Equation~\ref{eq:linear_connection} (also in Equation~\ref{eq:estimated_zeroth_moment} and ~\ref{eq:combo_zeroth_firt_moment}) that some residual leakage from nuisance components and instrument noise contaminates both estimated moments. Therefore, to propagate the uncertainties in both dimensions in the spectral index estimation, we adopt the `{effective variance
method}' \citep{Orear:1982,Petrolini:2014} that supports the standard deviation in both variables, minimising the error function,
\begin{equation}\label{eq:chi_sq_effective_variance}
    \chi^{2} = \sum_{p \in\ \mathcal{S}(p)} \frac{(m_1(p)-\hat{\beta}_s m_0(p))^2}{\sigma_1(p)^{2} + (\partial m_1(p)/\partial m_0(p))^2\sigma_0(p)^2},
\end{equation}
where $\sigma_0$(p) and $\sigma_1$(p) are the vector of noise standard deviation corresponding to $ n_0(p)$ and $n_1(p)$ respectively.
The maximum contribution of thermal dust and CMB is at \wmap\ Ka-band. Their typical level is 1-2\% of that of synchrotron. At channels having frequencies lower than \wmap\ Ka-band, this level is several times smaller. In the recovered maps, the dominant source of total residual leakage is instrumental noise. Consequently, when estimating the standard deviation vectors $\sigma_0(p)$ and $\sigma_1(p)$ in Equation~\ref{eq:chi_sq_effective_variance}, we account only for the contribution from noise. Specifically, we compute these vectors from multiple noise residual maps obtained by applying the same optimised weights to 100 independent noise realisations. The contribution from residual leakage of other nuisance components is therefore neglected. 

In order to minimise the systematic uncertainties related to the orientation angle used to pixelise the Stokes parameters with respect to the Galactic coordinate system, following \cite{Fuskeland:2021}, we marginalise over polarisation angles. That is, we perform the \textit{T-T} plot for Stokes parameters at a new coordinate system by rotating the plane of polarisation by an angle $\alpha$ that ranges from 0$^{\circ}$  to 85$^{\circ}$ in bins of 5$^{\circ}$. The final inferred spectral index is the inverse-variance-weighted mean of spectral index for all $\alpha$, i.e., $ \hat{\beta}_s = \sum_{\alpha}\frac{{\hat{\beta}_{s,\alpha}}/\sigma^2_{\beta_{s,\alpha}}}{1/\sigma^2_{\beta_{s,\alpha}}}$ and the error bar is the quadratic sum of systematic uncertainty, $[\max\,{\hat{\beta}_{s,\alpha}}-\min\,{\hat{\beta}_{s,\alpha}}]/2$ and  statistical uncertainty \citep{Watts:2024}. The minimum uncertainty found for different $\alpha$ rotations is considered to be the statistical uncertainty at respective super-pixels.

\section{Sky simulation}\label{sec:sims}
The polarised microwave sky is primarily composed of CMB, synchrotron, and thermal dust emission. The polarisation fraction of AME is limited below 1\% as suggested in various observational limits that are discussed in Section~\ref{sec:intro}. We simulate these components using \pysm\ \citep{pysm:2017,pysm3:2021} at \Nside\ = 64 and expressed at Rayleigh–Jeans ($\mu$K$_{RJ}$) units. In simulations, we use the bandpass to be a delta function for all channels. In addition to these sky components, the data also contains the instrument noise. Therefore, we add random realisation of noise maps to the sky simulations at respective channels. The simulation of noise realisation maps for different experiments is discussed in Section~\ref{sec:data}. In the following sections, we discuss the model of different sky components used in our simulation. 
\subsection{CMB}\label{sec:cmb}
We use a Gaussian random realisation of the lensed CMB map simulated using a set of $C_{\ell}$s. The $C_{\ell}$s are generated from CAMB \citep{CAMB:2000} for best-fit cosmological parameters from \planck\ 2018 results \citep{planck-VI:2020}. We ignore the primordial $B$-mode polarisation in CMB map by setting the tensor-to-scalar ratio parameter $r =$ 0.

\subsection{Synchrotron}\label{sec:syncchrotron}
The spectral energy distribution (SED) of synchrotron is commonly described in $\mu$K$_{RJ}$ units by a power-law model \citep{rybicki1991radiative},
\begin{equation}\label{eq:power_law}
   \left(\substack{Q^{sync}_{\nu}\\U^{sync}_{\nu}} \right) = \left(\substack{Q_{\nu_s}\\U_{\nu_s}}\right) \left(\frac{\nu}{\nu_s}\right)^{\beta_s},
\end{equation}
where $Q_{\nu_s},U_{\nu_s}$ are the synchrotron polarisation templates at a reference frequency of $\nu_s$ and $\beta_s$ is the synchrotron spectral index. 
We use $\ensuremath{\tt s1}$  model of synchrotron of \pysm\ where the synchrotron $Q$ and $U$ maps obtained from WMAP 9-year 23 \GHz\ \citep{Bennett_2013} is used as templates at reference frequency $\nu_{s}$ = 23 \GHz. The synchrotron spectral index $\beta_s$ used in simulation varies across the sky. This $\beta_s$ map is obtained in \cite{Miville-Deschenes:2008} fitting the Haslam 408 \MHz\ and \wmap\ 23 \GHz\ map. The template maps are smoothed to 5$^{\circ}$  and small-scale information is added to the final template maps following the method discussed in \cite{pysm:2017}.

\subsection{Thermal dust}\label{sec:thermal_dust}
In the microwave frequency range, the thermal dust emission is well described by a MBB in $\mu$K$_{RJ}$ units,

\begin{equation}\label{eq:mbb}
    \left(\substack{Q^{dust}_{\nu}\\U^{dust}_{\nu}} \right) = \left(\substack{Q_{\nu_d}\\U_{\nu_d}}\right)\left(\frac{\nu}{\nu_d}\right)^{\beta_d+1} \frac{e^{\gamma\nu_d}-1}{e^{\gamma\nu}-1},
\end{equation}
where $Q_{\nu_d},U_{\nu_d}$ are the dust polarisation templates at reference frequency $\nu_d$, $\beta_d$ is the dust spectral index and $\gamma = \frac{h}{K_{B}T_D}$ and $T_d$ is the dust temperature. 
We use one component dust model,  $\ensuremath{\tt d1}$ of \pysm. The model uses the dust polarisation maps obtained from \commander\ as a template at reference frequency $\nu_{d}$ = 353 \GHz\ \citep{Planck-X:2016}. The template maps are smoothed to FWHM = 2.6$^{\circ}$  and small-scale information is added following the method discussed in \cite{pysm:2017}. The spatially varying dust temperature $T_d$ and spectral index $\beta_d$ used in this model are obtained from \commander. 
\begin{figure}
    \centering
    \includegraphics[width=\linewidth]{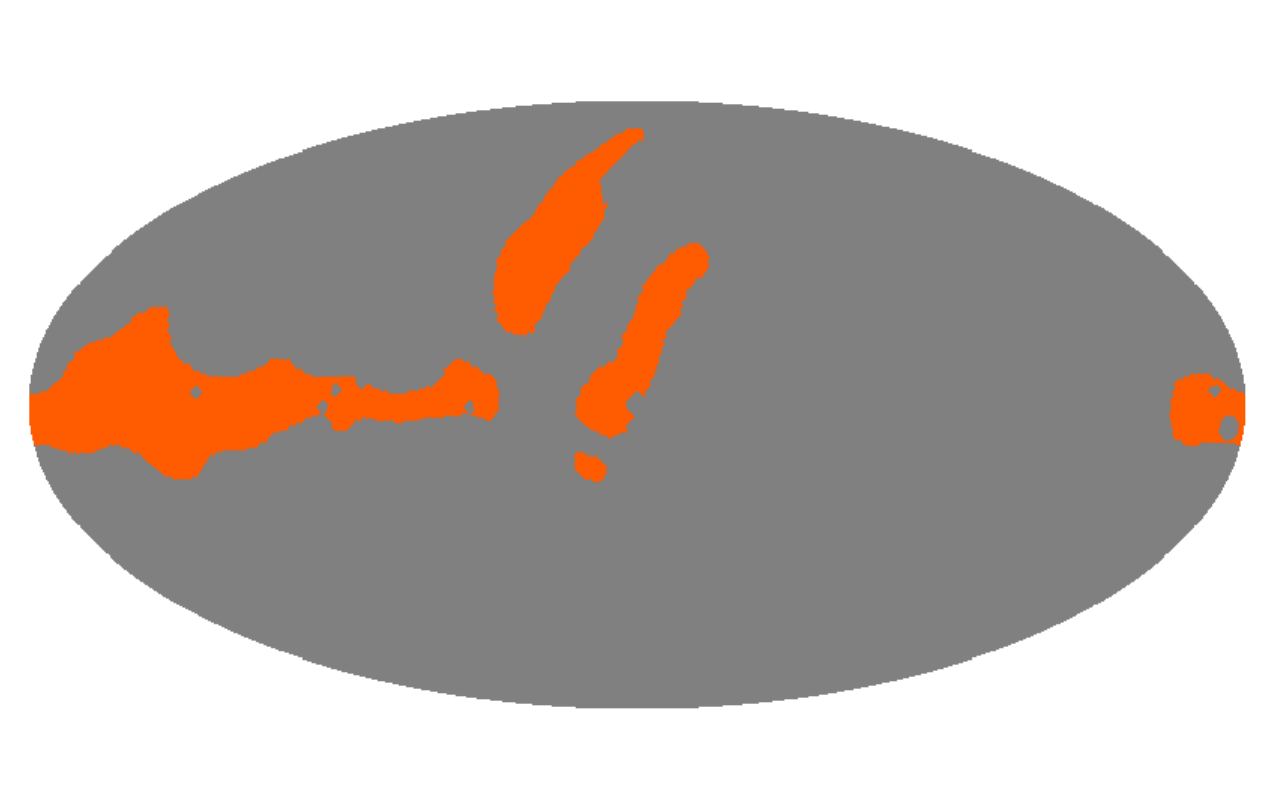}
    \caption{Mask at \Nside\ = 64 used in this analysis that retains 30\% high S/N regions of the QUIJOTE MFI survey. The bright source positions are masked using P06 polarisation mask provided by WMAP.}
    \label{fig:high_snr_mask}
\end{figure}
\subsection{AME}\label{sec:ame}
We simulate the AME maps from $\ensuremath{\tt a2}$ AME model of \pysm,
\begin{equation}\label{eq:ame}
    \left(\substack{Q^{AME}_{\nu}\\U^{AME}_{\nu}} \right) = p^{AME}I_{\nu}^{AME}\left(\substack{\cos(2\gamma_d)\\ \sin(2\gamma_{d})}\right),
\end{equation}
where $p^{AME}$ is the polarisation fraction. We modify the \pysm\ code to set $p^{AME}$ to 0.1\% and 0.5\% as motivated in section~\ref{sec:intro}. The AME intensity $I_{\nu}^{AME}$ has two components which follow two different emissivities that are obtained from the SpDust2 code \citep{Ali:2009}. The templates of spatially varying two components are obtained from \commander\ fitting \citep{Planck-X:2016}. The AME intensity maps are translated to polarisation maps using the same dust polarisation angle $\gamma_d$ obtained from $Q, \,U$ dust maps of \commander\ fitting.

\section{Data}\label{sec:data}
We use frequency maps below 40 \GHz\ which can be treated as synchrotron tracers.  In this paper, the data products we use are from the following three surveys.\\  

(i) QUIJOTE: The QUIJOTE (Q-U-I JOint TEnerife) MFI survey\footnote{\url{https://lambda.gsfc.nasa.gov/product/quijote/}} \citep{Rubiño-Martín:2023} has observed approximately 70\% of the sky across four frequency bands: 11, 13, 17, and 19 GHz. Among the four channels, the 11 GHz and 13 GHz channels provide the highest S/N to synchrotron emission. However, the noise at 13 GHz is observed to be  $\sim$30\% correlated with that at 11 GHz \citep{Rubiño-Martín:2023} for polarisation. Therefore, to maximise the effective S/N, we obtain our main results in our analysis using a weighted combined map of the 11 GHz and 13 GHz channel data. We assign weights of 55\% and 45\% respectively for 11 and 13 \GHz\ channels based on their sensitivity. In addition to this baseline data combination, we also repeat the full analysis using the 11 GHz and 13 GHz maps individually to assess the impact of each channel on the inferred results. Certain regions of the survey suffer from reduced sensitivity due to factors such as the presence of geostationary satellites and increased atmospheric contamination in specific directions. The S/N is particularly high near the Galactic plane and the North Polar Spur (NPS) regions. To ensure the robustness of our analysis, we restrict ourselves to the high S/N regions that encompass 30\% of the sky of the QUIJOTE MFI survey area. This selected area includes prominent features such as the Fan region and the NPS extending north Galactic plane and near the Galactic centre. Additionally, following \cite{Watts:2024}, we apply the P06 polarisation mask provided by WMAP\footnote{\url{https://lambda.gsfc.nasa.gov/product/wmap/dr2/masks_info.html}} to exclude the brightest emission regions. The final mask used in our analysis is shown in Figure~\ref{fig:high_snr_mask}.\\  

(ii) WMAP: We consider WMAP 9-year data\footnote{\url{https://lambda.gsfc.nasa.gov/product/wmap/dr5/m_products.html}} at K and Ka bands \citep{Bennett_2013}. We do not consider other bands of \wmap\ data due to the lower S/N of the synchrotron signal. \\

(iii) \planck: We use \planck\ low-frequency instrument (LFI) data at 30 \GHz\ from PR3 data release\footnote{\url{https://pla.esac.esa.int}} \citep{Planck-I:2020, Planck-II:2020}. We verified in Appendix~\ref{sec:pr4_results} whether the recovered spectral index remains essentially unchanged when using PR4 \citep{npipe:2020} instead.

An important consideration in the use of QUIJOTE MFI data is the possible impact of Faraday rotation on synchrotron emission \citep{beck_R:2013}. Faraday rotation, which alters the polarisation angle of incoming radiation, becomes increasingly significant at lower frequencies due to its dependence on the square of the wavelength \citep{beck_R:2013}. This effect can lead to a flattening of the synchrotron spectral index, particularly near the Galactic plane, where the electron density and magnetic fields are stronger. To assess the potential influence of Faraday rotation in our analysis, we examined the spatial correlation between the Stokes 
$Q$ and  $U$ maps from WMAP K-band and QUIJOTE MFI 11 \GHz\ data. We found a high degree of correlation between the two frequency channels, which suggests the presence of a coherent magnetic field largely aligned with the Galactic plane. This also indicates that the level of Faraday depolarisation near the Galactic plane is negligible, which is our region of interest. Furthermore, since \cite{Rubiño-Martín:2023} reported in their analysis that there are no changes in the final inference after correcting for Faraday rotation using the rotation measure from \cite{Hutschenreuter:2020}. Therefore, we do not apply any correction for Faraday rotation in this analysis.

Before applying the cPLIC method, all data are processed as follows:\\
(i) The residuals of Radio Frequency Interference (RFI) mainly due to emissions from geostationary satellites
projected in the map-making process appear at fixed azimuth locations in \quijote\ MFI frequency maps. This residual RFI is subtracted using a function of the declination (FDEC\footnote{\url{https://github.com/jarubinomartin/sancho}}). This function applies a filter to the maps to remove the median value of all pixels at constant declination. \cite{de_la_Hoz:2023} reported that if we apply the same FDEC correction to the other data sets, the potential bias in $\beta_{s}$ gets reduced significantly. Therefore, and in order to maintain the consistency of zero levels to the other data sets, the same FDEC filter is applied to \wmap\ and \planck\ data.\\
(ii) Our analysis requires all the maps to have the same resolution and to be pixelised on the same \healpix\ grid \citep{Healpix, Healpix2}. 
Therefore, we apply the FDEC correction to all frequency maps at the common one degree resolution and \Nside\ = 512, and after this, we further convolve the maps to a final Gaussian beam of FWHM of 2$^{\circ}$ to increase the signal to noise.
Finally, all maps are downgraded to \Nside\ = 64 and converted to $\mu$K$_{RJ}$ units. 

We also use the noise realisation maps at these frequencies in order to estimate the variance of residual leakage to recovered synchrotron moment maps using the cPILC method. We consider \quijote\ MFI simulated noise maps that include the 1/\textit{f} noise \citep{Rubiño-Martín:2023}. We simulate the \wmap\ noise using Gaussian random realisations for standard deviation maps, $\sigma (p) = \sigma_{0}/\sqrt{N_{obs}(p)}$, where $\sigma_{0}$ is RMS noise at respective frequencies \citep{Hinshaw:2003} and $N_{obs}$ is the hit map of \wmap\ mission\footnote{Both are provided in \url{https://lambda.gsfc.nasa.gov/product/wmap/dr5/sk
ymap info.html}}. For \planck\ channels, we use noise maps of end-to-end noise simulation \citep{Planck-xii:2016}. All the noise realisation maps are also smoothed to a common beam resolution of FWHM = 2$^{\circ}$, downgraded to \Nside\ = 64 and converted to $\mu$K$_{RJ}$ units.   

\begin{figure}
    \centering
    \includegraphics[width=0.5\textwidth]{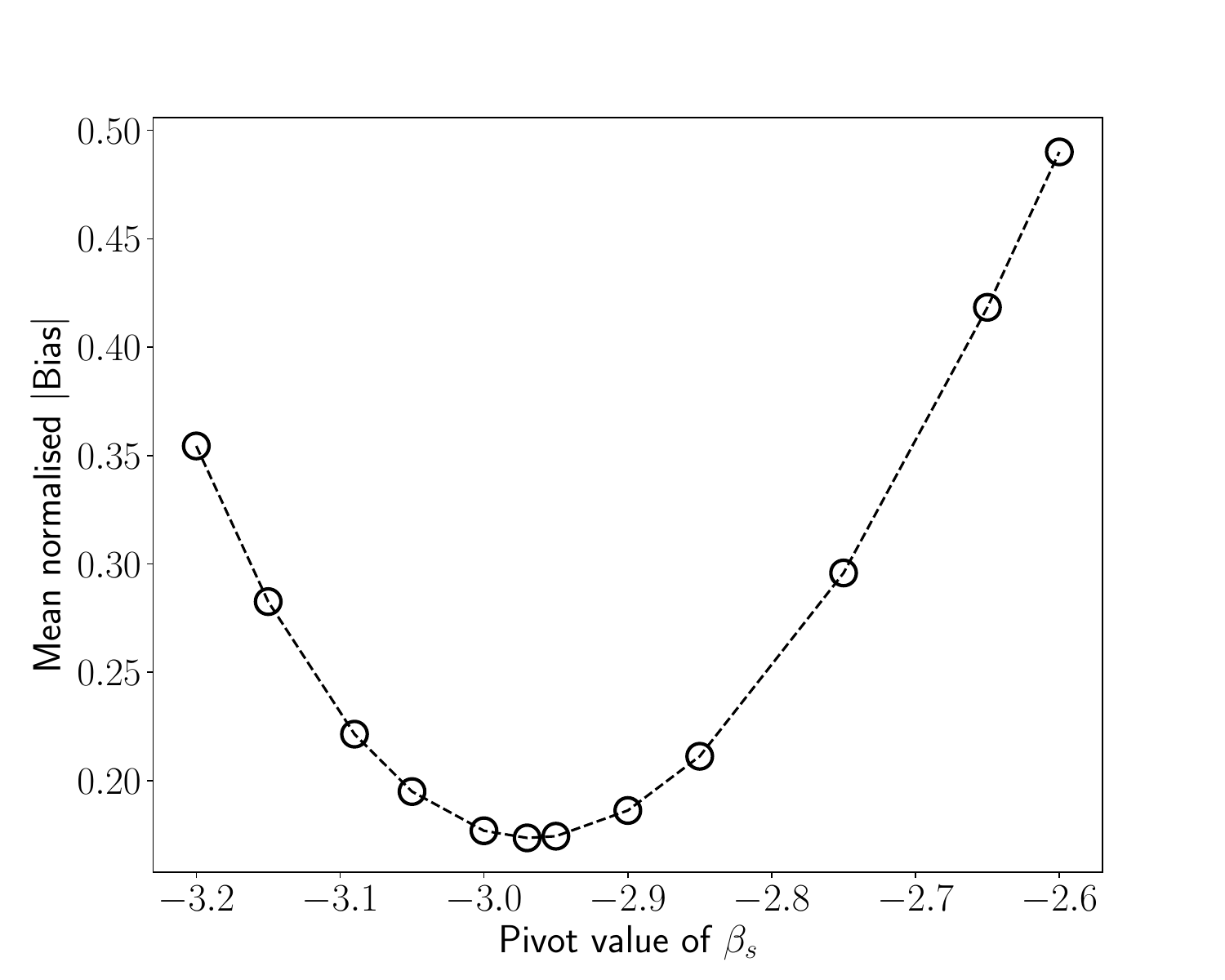}
    \caption{Changes of the mean absolute deviation of the inferred spectral index with respect to the input synchrotron spectral index template, normalised with respect to the mean error of estimated $\beta_s$ with a different choice of the pivot value of the synchrotron spectral index.}
    \label{fig:bias_wrt_choice_pivot}
\end{figure}

\section{Validation on simulated maps}\label{sec:sim_val}
 We first apply our method to two different sets of simulations for validation. We start with the simulation without a polarised AME component. We discuss the obtained results analysing this set of simulations in Section~\ref{subsec:recoverd_moments} and Section~\ref{subsec: betas_estimate}. To investigate the impact of AME polarisation, next, we incorporate a polarised AME component into the simulation. We discuss the corresponding results in Section~\ref{sec:ame_bias}. 
\begin{figure*}
    \includegraphics[width=0.5\textwidth]{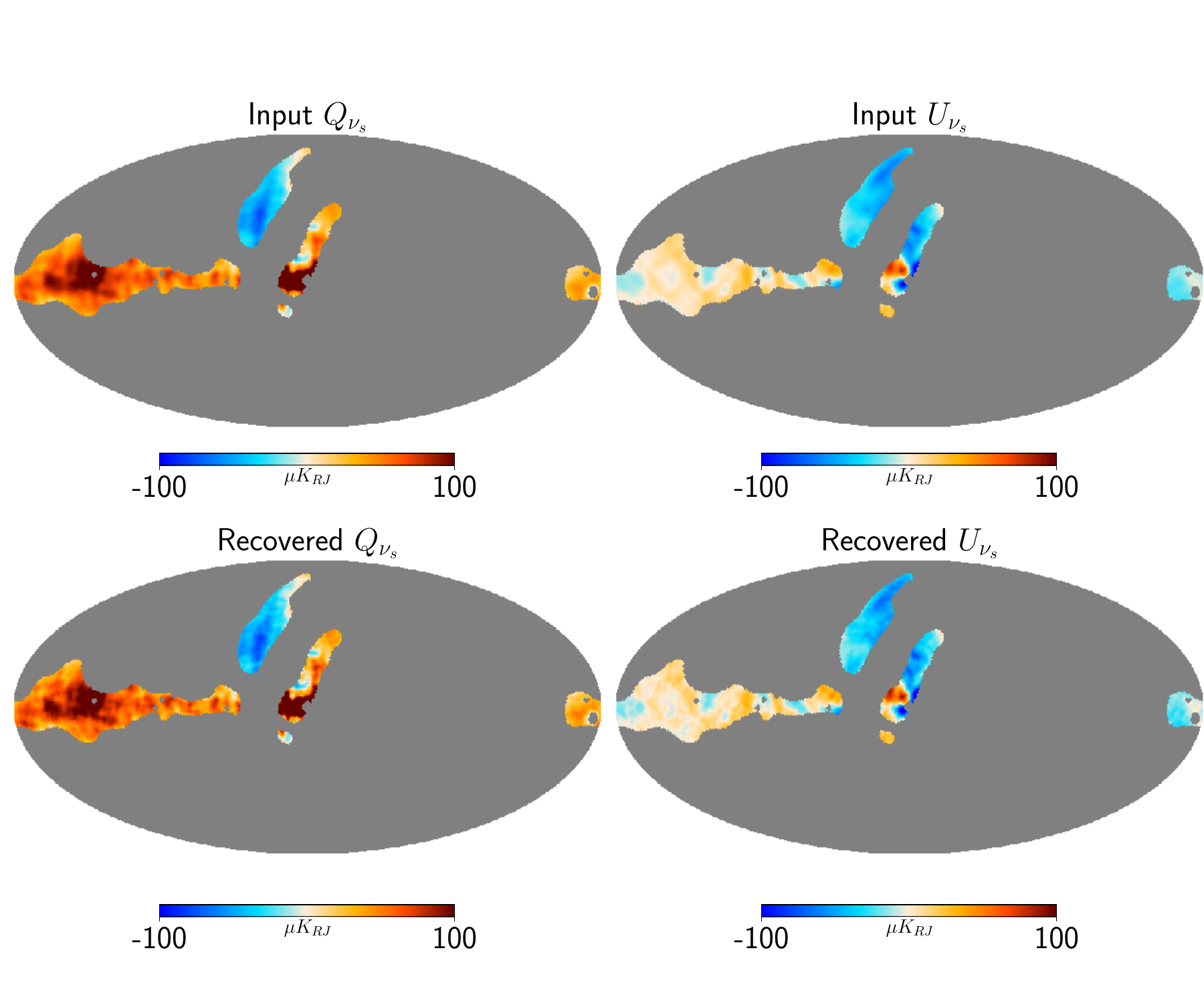}
    \includegraphics[width=0.5\textwidth]{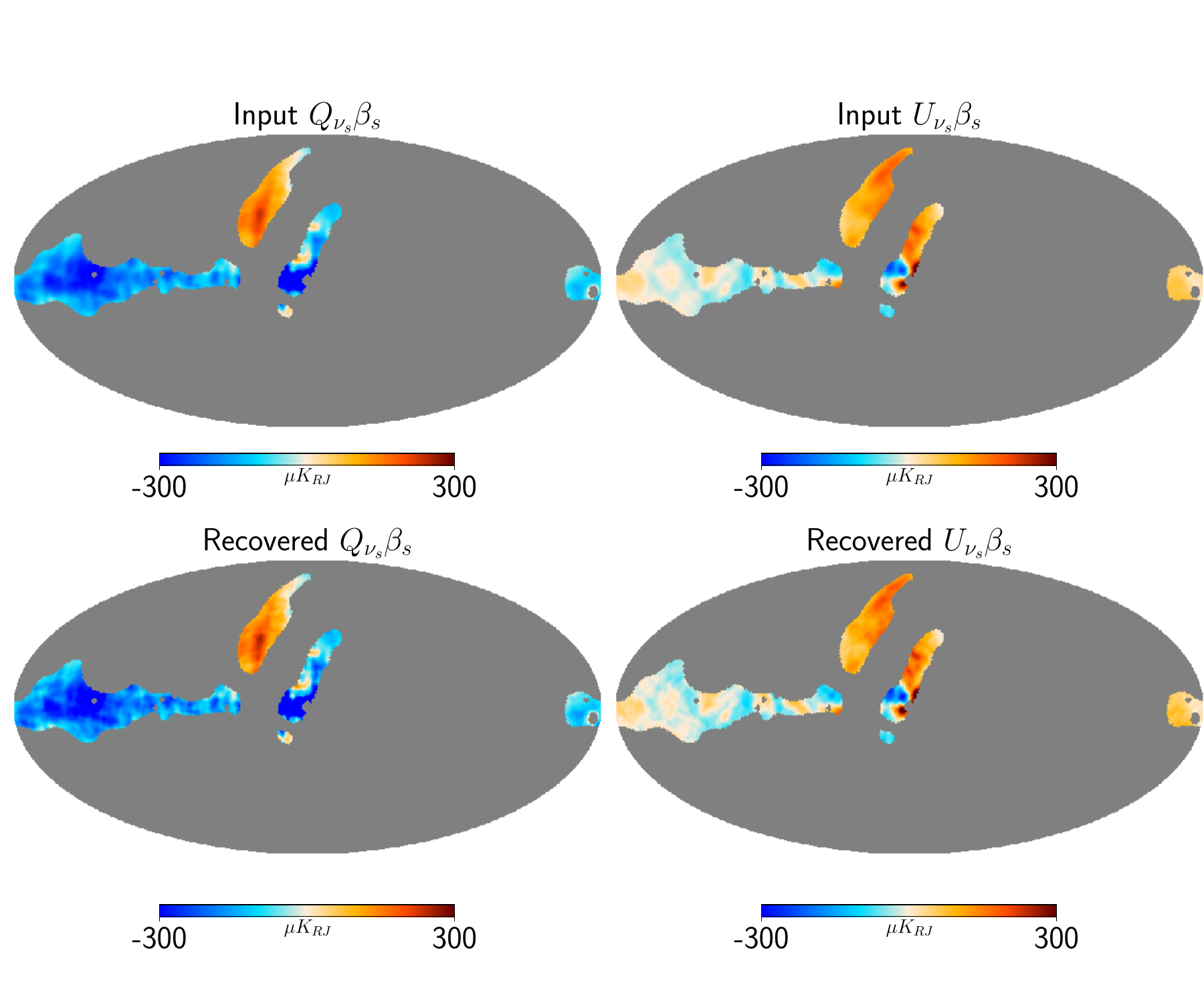}
    \caption{Comparison between input (upper panel) and recovered (lower panel) maps of $Q_{\nu_s}(p)$, $U_{\nu_s}(p)$ and $Q_{\nu_s}(p)\beta_s(p)$, $U_{\nu_s}(p)\beta_s(p)$ at the reference frequency of 23 \GHz. All maps are shown at resolution of beam FWHM = 
    2$^{\circ}$. The correlation between input and recovered maps is found to be 99\% for $Q$ and $95$\% for $U$. A similar correlation is found for $Q_{\nu_s}(p)\beta_s(p)$ and $U_{\nu_s}(p)\beta_s(p)$ between input and recovered maps.}
    \label{fig:compare_zeroth_moment}
\end{figure*}

\begin{figure}
    \centering
    \includegraphics[width=0.5\textwidth]{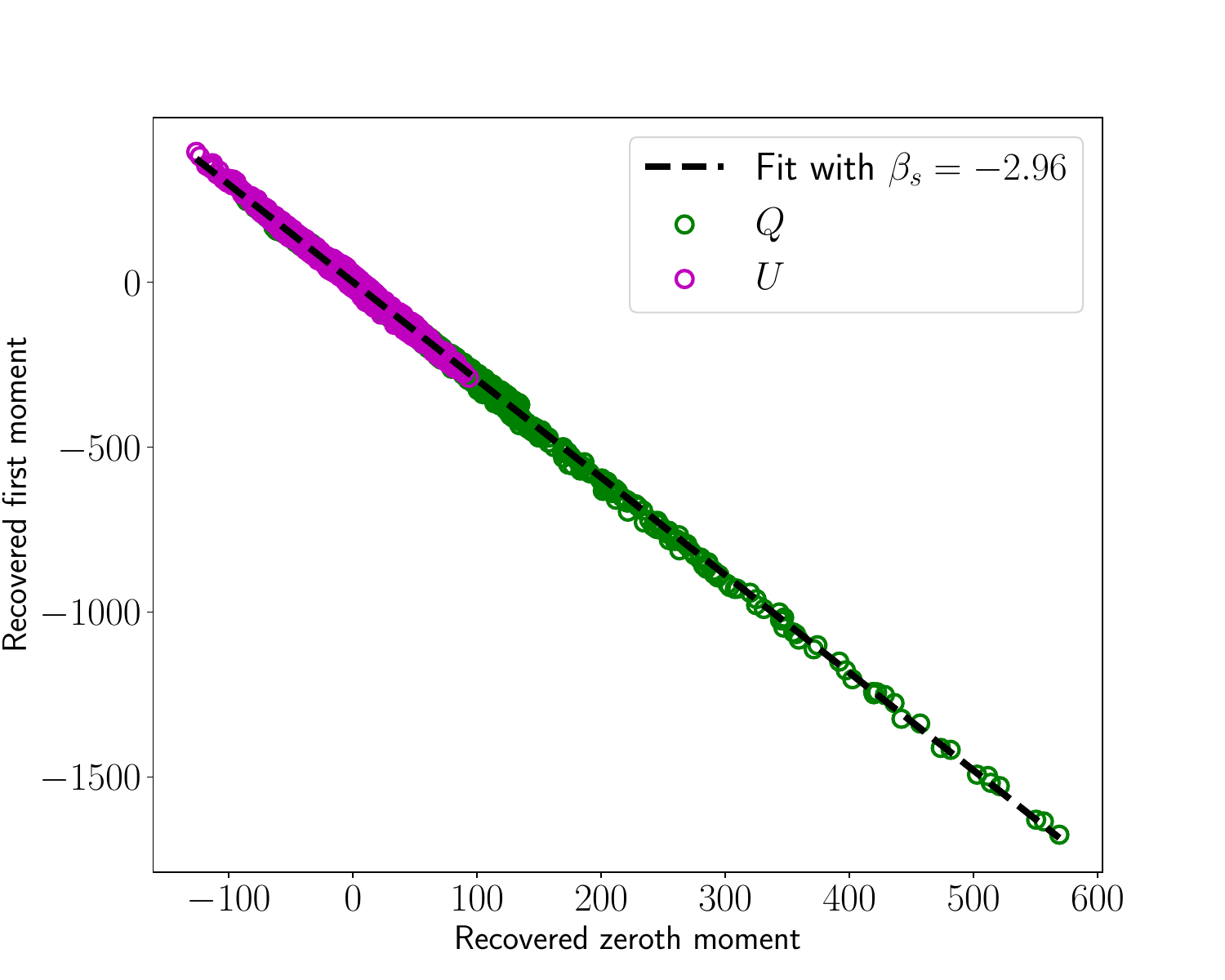}
     
    \caption{Correlation between recovered Stokes $Q(p)$ (green) and $U(p)$ (magenta) maps corresponding to zeroth and first order moments. The black dashed line is the linear fit with $\beta_s = -2.96$ which corresponds to mean value of the input spectral index map. }
    \label{fig:moment_correlation_pysm}
\end{figure}

\begin{figure*}
    \includegraphics[width=\textwidth]{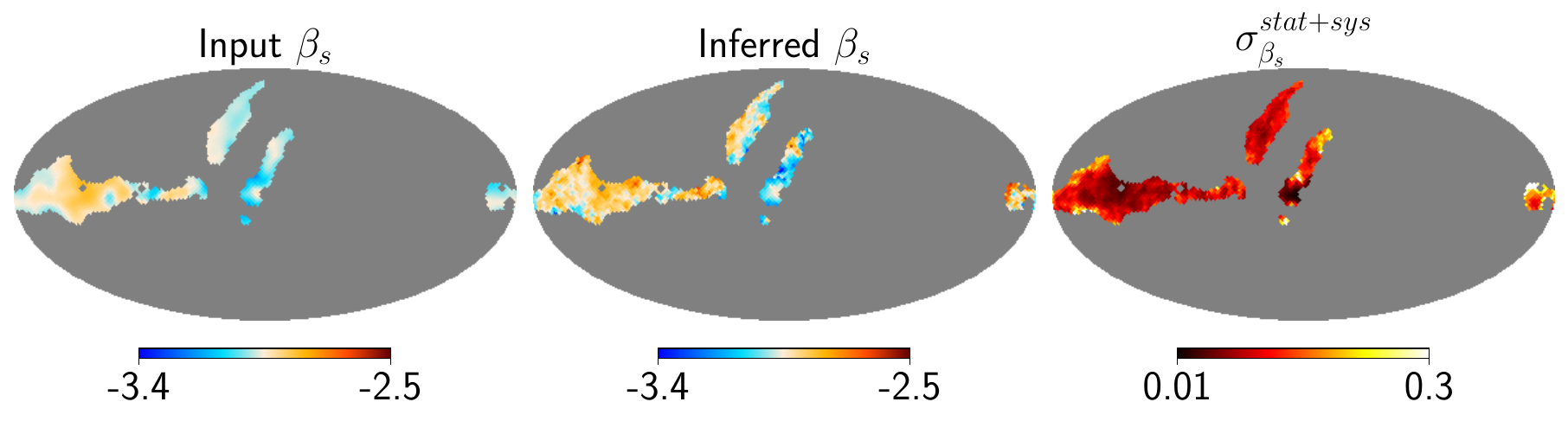}
\caption{Left panel: The input $\beta_s$ map used in the simulation. Middle panel: The inferred $\beta_s$ map from the estimated synchrotron moments. Right panel: The map of statistical and systematic uncertainty of the estimated $\beta_s$ which are added in quadrature.} 
    \label{fig:beta_input_estimated_mean_std_noamesims}
\end{figure*}

\begin{figure}
    \includegraphics[width=\linewidth]{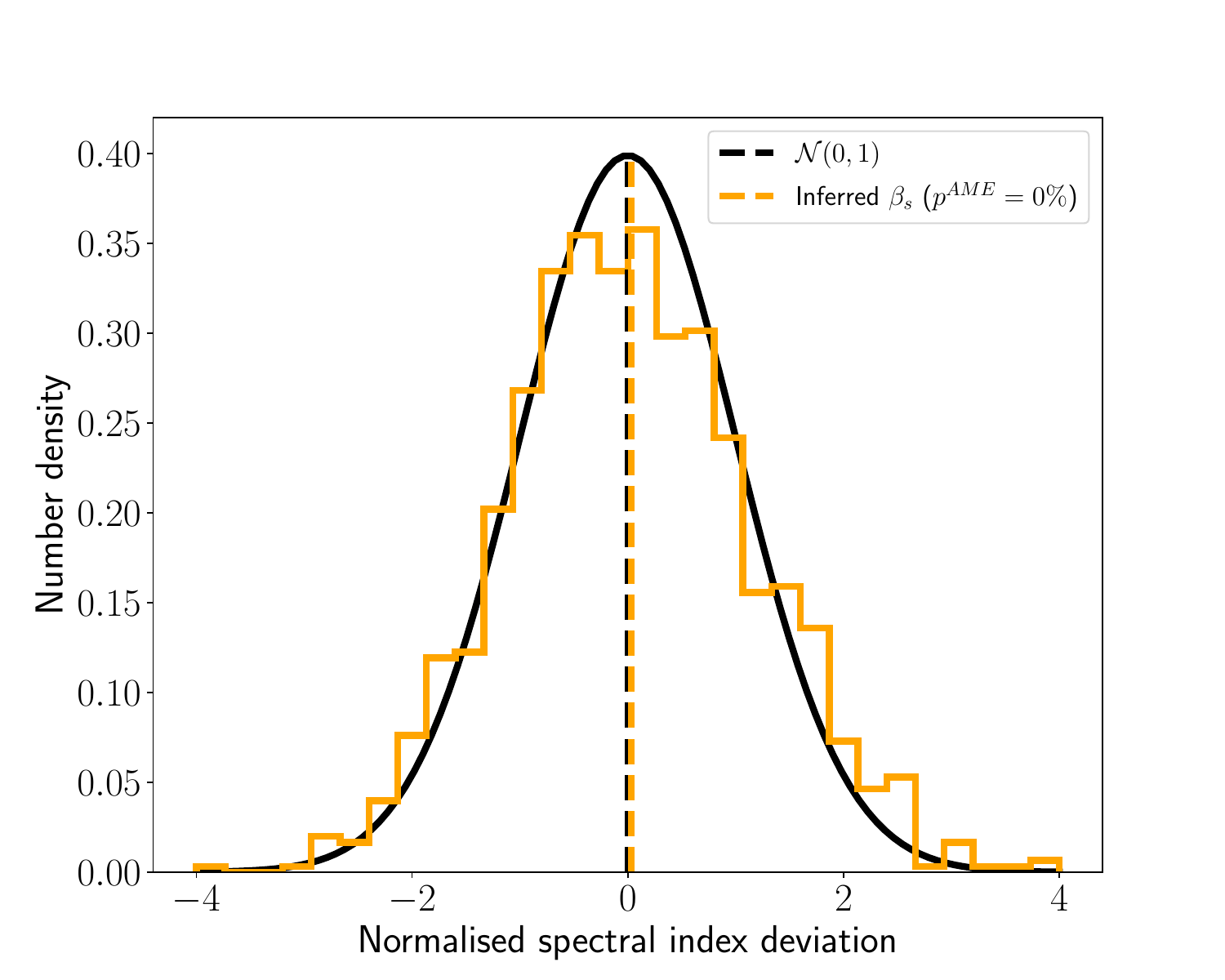}
    \includegraphics[width=\linewidth]{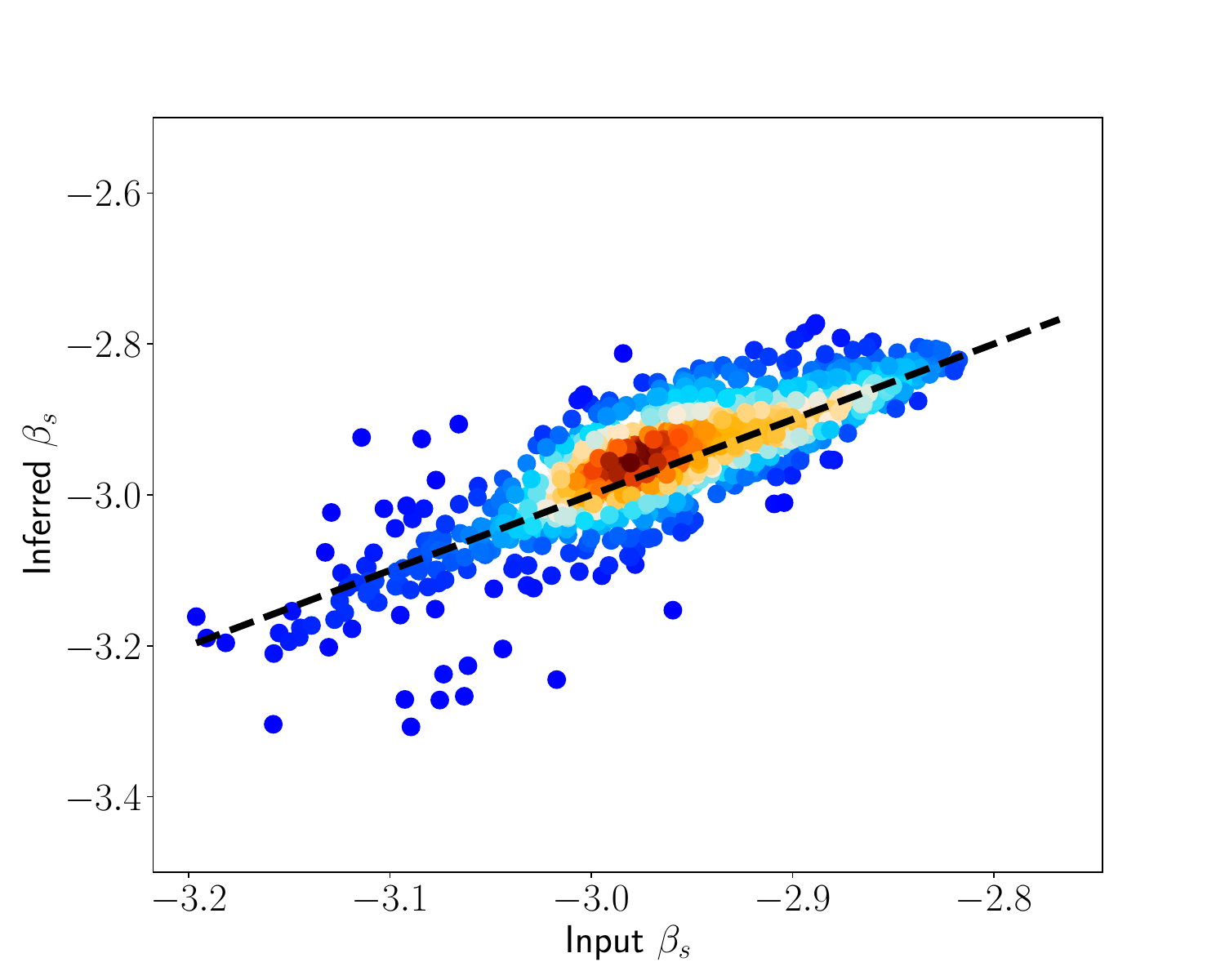}
    
   \caption{Upper panel: Histogram of the normalised deviation of the derived spectral index with respect to the true input spectral index $(\beta_s^{\rm Inferred} - \beta_s^{\rm Input})/\sigma_{\beta_s}^{\rm stat+sys}$ (orange). The black curve is a Gaussian distribution with zero mean and unit variance. Vertical lines are the corresponding mean of the distribution. Lower panel: The two-dimensional density plot of true input and inferred  $\beta_s$ map. The correlation between the two maps is 81\%. The black dashed line correspond to 100\% correlation. }
    \label{fig:beta_bias_std_noamesims}
\end{figure}

\begin{figure}
    \includegraphics[width=\linewidth]{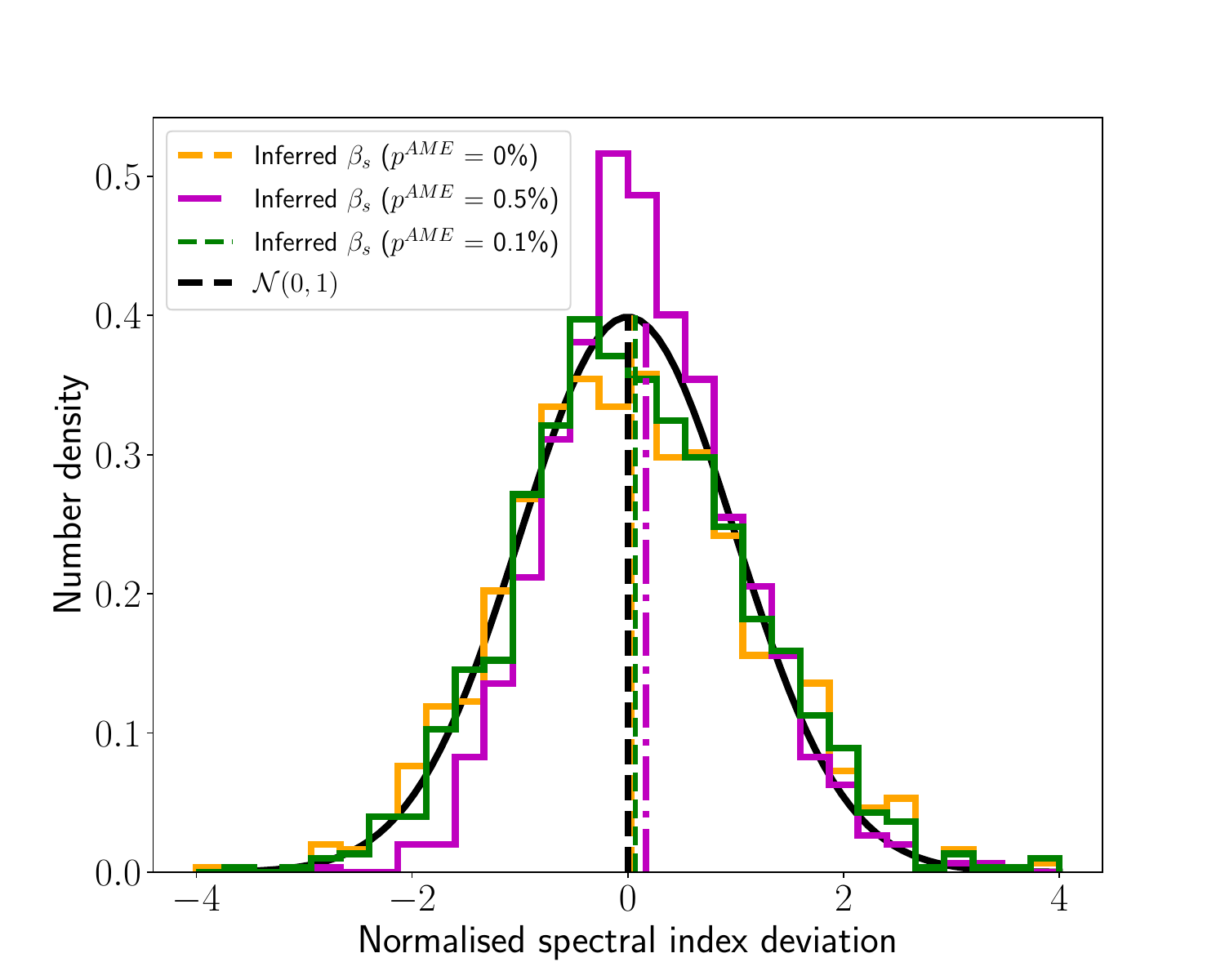}%
   \caption{Histograms of the normalised deviation of the spectral index derived from simulations of three cases: without polarised AME (orange), with 0.5\% (magenta), and 0.1\% (green) polarised AME. The black curve is a  unit variance Gaussian distribution with zero mean. Vertical lines are the corresponding mean of the distributions.}
    \label{fig:beta_bias_std_amesims}
\end{figure}

\begin{figure*}
    \includegraphics[width=\linewidth]{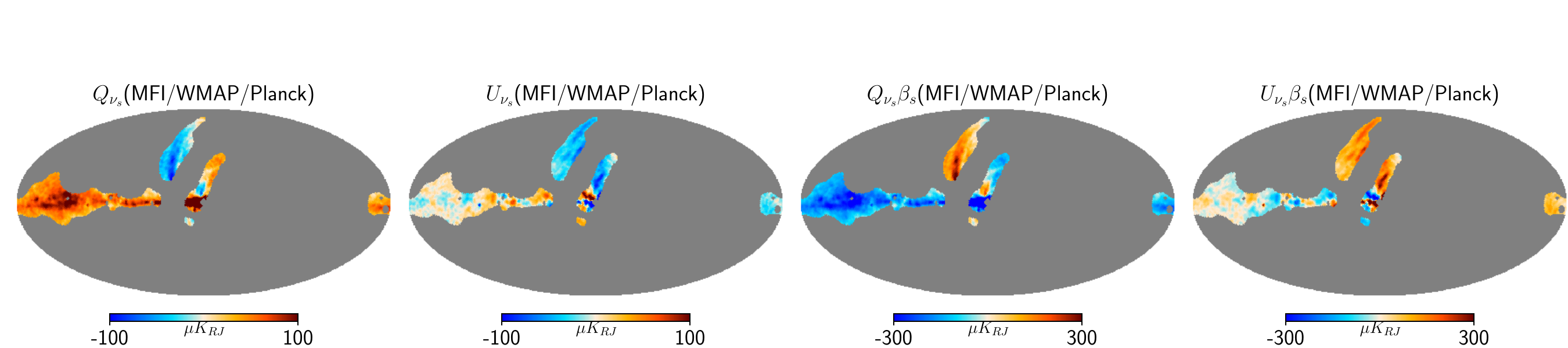}
    
   \caption{Synchrotron Stokes $Q_{\nu_s}$ (first panel) and $U_{\nu_s}$ (second panel) maps at 22.8 \GHz\ recovered from the cPILC component separation method using the QUIJOTE MFI 11 and 13 \GHz\ combined map, WMAP K and Ka bands and \planck\ LFI 30\GHz. The Stokes parameters corresponding to the spectral index modulated synchrotron amplitudes, $Q_{\nu_s}\beta_s,U_{\nu_s}\beta_s$ obtained from component separation are shown in  the third and fourth panels respectively. All maps are at FWHM=2$^\circ$ at \Nside = 64.  }
    \label{fig:moment_maps_data}
\end{figure*}

\subsection{Reconstruction of synchrotron moment maps}\label{subsec:recoverd_moments}
We apply the cPILC method to simulated data to recover the synchrotron moment maps while cleaning the nuisance components described in Equation~\ref{eq:data_representation}. Specifically, we obtain maps of the synchrotron Stokes parameters, $Q_{\nu_s}(p)$ and $U_{\nu_s}(p)$, as well as the spectral index modulated Stokes parameters, $Q_{\nu_s}(p)\beta_s(p)$ and $U_{\nu_s}(p)\beta_s(p)$, at a reference frequency of $\nu_s = 22.8$~\GHz, following Equations~\ref{eq:estimated_firts_moment} and \ref{eq:combo_zeroth_firt_moment}, respectively. These two sets of maps have zero contamination from one to another. To minimise the contribution from synchrotron higher-order moments to residual leakage, we carefully select the pivot value of the synchrotron spectral index. This is achieved by iteratively varying the pivot value within the range of $-3.2$ to $-2.6$. Figure~\ref{fig:bias_wrt_choice_pivot} shows the mean absolute difference between the inferred and input $\beta_s$, normalised to the corresponding mean uncertainty $\sigma_{\beta_s}$, as a function of the chosen pivot value. We find that the deviation of the inferred spectral index is smallest when the pivot value is close to the central value of the inferred $\beta_s$ distribution. The deviation remains nearly unchanged for pivot values between $-3.0$ and $-2.9$. Therefore, we adopt an iterative scheme that updates the pivot until the inferred distribution stabilises.  Without deprojecting thermal dust emission, we find that the recovered synchrotron moments; particularly the first-order moment maps contain noticeable residuals from thermal dust. To mitigate this, we impose an additional constraint as described in Equation~\ref{eq:constraints_on_weights_c} to deproject the zeroth moment of thermal dust. We adopt the \planck\ best-fit values of thermal dust spectral parameters at $\overline{T}_d = 19.8 $K and $\overline{\beta}_d = 1.55$ \citep{Planck-IV:2020} to be the pivot values of the MBB parameters while deprojecting the dust moment. In Figure~\ref{fig:compare_zeroth_moment}, we compare the input (upper panel) and cPILC-recovered (lower panel) maps of $Q_{\nu_s}(p),U_{\nu_s}(p)$ and $Q_{\nu_s}(p)\beta_s(p),U_{\nu_s}(p)\beta_s(p)$ at the reference frequency of 22.8 \GHz. All the maps are displayed at a resolution of 2$^{\circ}$. The method enables us to recover both the moments with high fidelity, as evidenced by the strong spatial correlation between the input and recovered maps. The recovered $Q$ maps appear cleaner than the $U$ maps, reflecting the fact that the variance minimisation in the cPILC method is dominated by the stronger $Q$ signal. Finally, we examine the correlation between the two recovered moment maps. Figure~\ref{fig:moment_correlation_pysm} shows the T–T correlation between them, yielding correlation coefficients of 0.99 for Stokes $Q$ and 0.95 for Stokes $U$. The black dashed line indicates a linear fit with $\beta_s = -2.96$, corresponding to the mean value of the synchrotron spectral index template used in the simulation.
 
\subsection{Inference of synchrotron spectral index }\label{subsec: betas_estimate}
The two synchrotron moment maps recovered in the previous section are found to be anti-correlated. We exploit this anti-correlation to infer the synchrotron spectral index, $\beta_s$, by performing a linear regression (“T–T plot”) between the two derived synchrotron moment maps, as described in Section~\ref{sec:TT_correlation}. The regression is carried out over sub-pixels at \Nside = 64 contained within each super-pixel at \Nside = 32. We estimate the $\beta_s$ for 18 rotation angles within 0$^\circ$ to 85$^\circ$ and then compute the inverse-variance-weighted mean of the spectral index from measurements of all rotation angles. The resulting map of spectral index at \Nside=32 is presented in the middle panel of Figure~\ref{fig:beta_input_estimated_mean_std_noamesims}. We present the input reference $\beta_s$ map in the left panel of this figure for comparison. The right panel presents the corresponding uncertainty map, $\sigma_{\beta_s}^{\mathrm{stat+sys}}$, which combines both statistical and systematic errors (stat + sys) in quadrature, as discussed in Section~\ref{sec:TT_correlation}. Visually, the recovered spectral index map closely resembles the input  $\beta_s$ map. However, a small level of disagreement is expected. The \pysm\ pipeline \citep{pysm:2017} uses the synchrotron spectral index template from \citet{Miville-Deschenes:2008}, smoothed to $5^{\circ}$. Additional small-scale fluctuations in the simulated frequency maps are introduced later through the power spectrum, as outlined in Section~\ref{sec:syncchrotron}. The final simulated maps are smoothed to $2^{\circ}$ in our post-processing before applying the cPILC method. These non-linear processes naturally lead to minor discrepancies between the recovered and input spectral index maps. For a more quantitative assessment, the upper panel of Figure~\ref{fig:beta_bias_std_noamesims} shows the histogram of the normalised deviation of the inferred spectral index, $(\beta_s^{\mathrm{Inferred}} - \beta_s^{\mathrm{Input}})/\sigma_{\beta_s}^{\mathrm{stat+sys}}$ (orange). If the estimation is unbiased and the uncertainties are well characterised, this distribution should follow a normal function, $\mathcal{N}(0,1)$, represented by the solid black line. We find that, across a large fraction of the analysed region, the inferred $\beta_s$ values are consistent with the input within $3\sigma$ confidence. Only about 13\% of our analysis region exhibits lower confidence, primarily due to the reduced S/N in those regions.
To further evaluate the performance, we examine the spatial correlation between the inferred and input $\beta_s$ maps. The lower panel of Figure~\ref{fig:beta_bias_std_noamesims} presents the pixel-by-pixel correlation between the two. 
We find a strong correlation of 81\% between the inferred and input $\beta_s$ maps. Within the analysed region, the input $\beta_s$ distribution has a mean of $-2.96$ and a standard deviation due to intrinsic scatter of $0.06$, while the corresponding inferred distribution yields a mean of --2.97 and a standard deviation of $0.16$.

\subsection{Spectral index inference in presence of polarised AME}\label{sec:ame_bias}
The polarisation fraction of diffuse AME remains relatively uncertain. As noted in Section~\ref{sec:intro}, several previous theoretical and empirical studies have suggested that AME can be polarised at level below 1\% \citep{Drain_hensley:2013,Hoang_alex:2016,Génova-Santos:2017,Gonz:2024}. In this section, we investigate the impact of polarised diffuse AME on the estimation of the synchrotron spectral index. We perform our analysis using two sets of simulations based on the \pysm\ $\tt a2$ model, incorporating polarised AME with polarisation fractions of 0.5\% and 0.1\%, respectively. It is important to emphasise that we do not impose any extra constraint in our cPILC method to deproject the AME moments while recover the synchrotron moments, which would increase the residual noise  level in the recovered maps and lead to greater dispersion in the inferred spectral index. Figure~\ref{fig:beta_bias_std_amesims} illustrates the comparison of the normalised spectral index deviations for three cases: simulations without polarised AME (orange), with 0.5\% polarised AME (magenta), and with 0.1\% polarised AME (green). Our results indicate that the presence of a 0.5\% polarised AME component causes the inferred synchrotron spectral index to appear slightly flatter compared to the unpolarised case. The mean and standard deviation of the spectral index in this scenario are --2.96 and 0.16, respectively, indicating that the mean value is shifted by only 0.5\% relative to the unpolarised AME results. The $\beta_s$ values inferred from the polarised and unpolarised AME cases are consistent within 1.5$\sigma$ across all pixels.
When the AME polarisation fraction is reduced to 0.1\%, the mean and standard deviation of the inferred spectral index become --2.95 and $0.15$, respectively, which are statistically indistinguishable from the unpolarised AME case, showing no measurable bias.

\section{Results on real data}\label{sec:results}
We now proceed to apply the component separation method to the 2$^{\circ}$ smoothed maps at frequencies < 40 \GHz\ from \planck\ PR3 LFI 30 \GHz, WMAP K and Ka bands  and QUIJOTE MFI 11 and 13 \GHz\ channels. The results obtained with \planck\ PR4 data at 30 \GHz\ are presented in Appendix~\ref{sec:pr4_results}. For the QUIJOTE MFI data, the 11 \GHz\ and 13 \GHz\ channels are combined using weights of 55\% and 45\%, respectively, to avoid accounting for the correlated noise between two channels as stated in Section~\ref{sec:data}. This choice of the weights are not strictly optimal, instead they are chosen to take into account the higher S/N at 11 \GHz\ as compared to 13 \GHz.  In addition to this analysis with combined MFI maps, we have also used the 11 \GHz\ and 13 \GHz\ maps separately in cPILC algorithm and perform subsequent analysis. The corresponding results are discussed later in this section. The constraint equations given in Eqs.~\ref{eq:constraints_on_weights_a}, \ref{eq:constraints_on_weights_b}, and \ref{eq:constraints_on_weights_c} are imposed on the effective spectral response of each component, computed integrating the instrument-specific bandpass profiles \citep{Jarosik:2003,PLANCK-V:2014,Rubiño-Martín:2023} over the spectral behaviour of respective components. Given that the WMAP 9-year map-making process incorporates bandpass shift, we adopt a similar treatment of bandpass variation in our analysis, as detailed in \cite{Bennett_2013}. For the combined QUIJOTE MFI data, we applied the same channel weights in the bandpass integration as those used to produce the combined maps. From the component-separated maps, we reconstruct the Stokes parameter maps corresponding to synchrotron amplitude and spectral index modulated synchrotron amplitude following Eq~\ref{eq:estimated_zeroth_moment} and Eq~\ref{eq:estimated_firts_moment} at \Nside = 64. We display the resulting maps in Figure~\ref{fig:moment_maps_data}. The figure demonstrates a clear spatial anti-correlation between the zeroth and first order recovered moment maps, a feature that is further utilised in the estimation of the synchrotron spectral index. 

\begin{figure}
    \centering
    \includegraphics[width=0.5\textwidth]{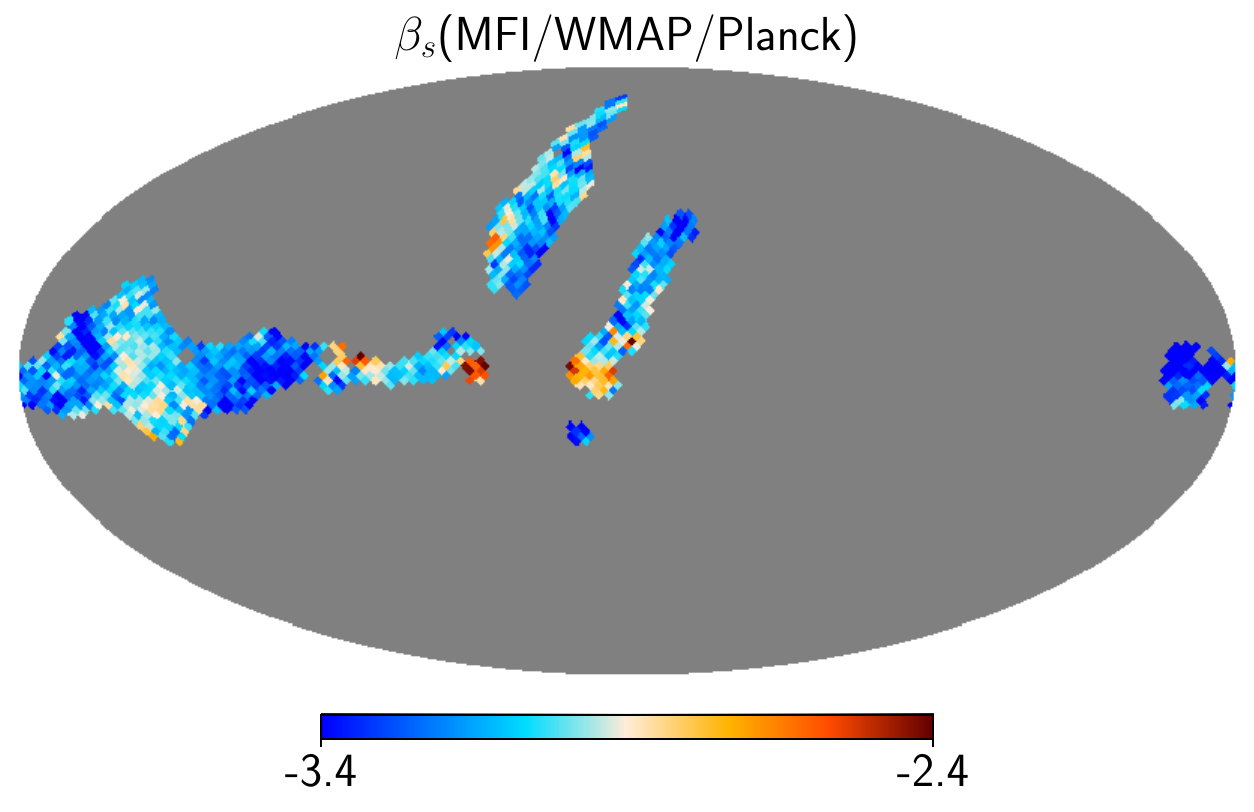}
    \includegraphics[width=0.5\textwidth]{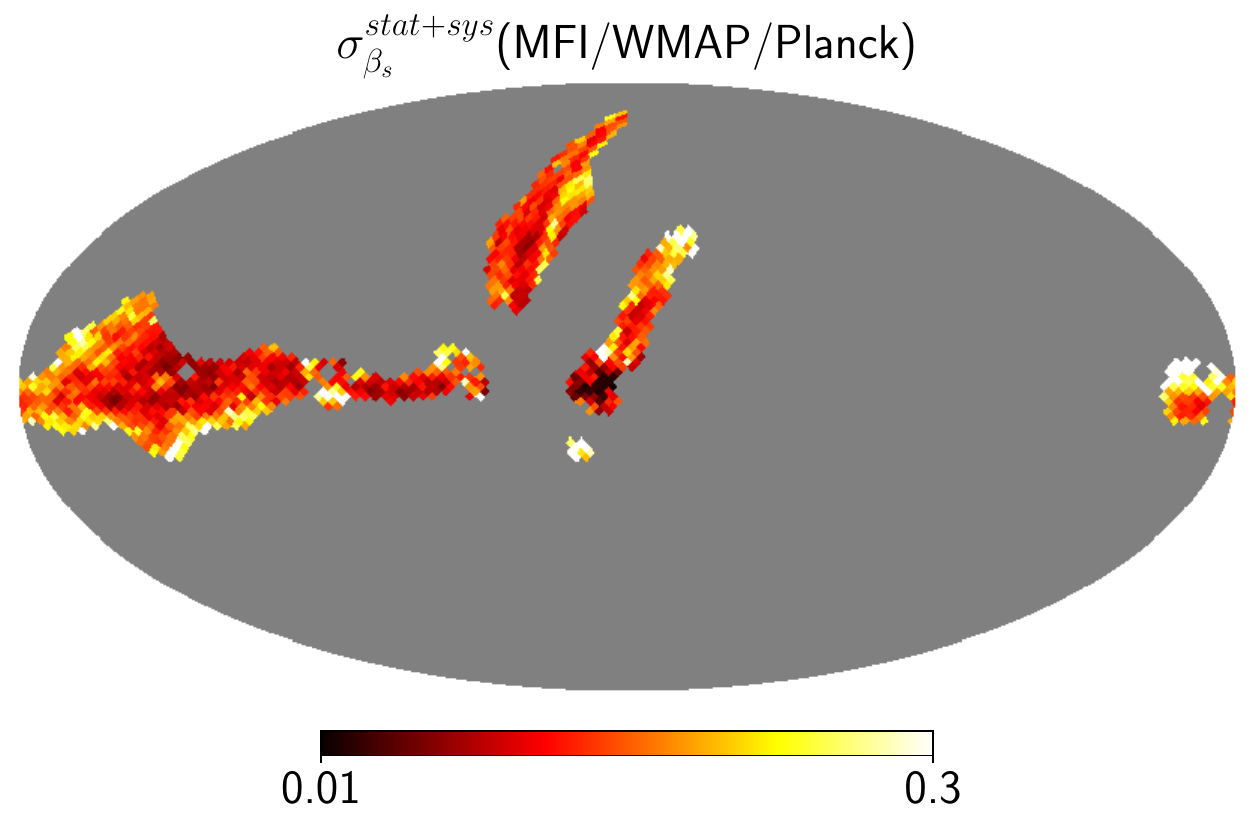}
    \caption{ Synchrotron spectral index (upper panel) and uncertainty maps (lower panel) obtained from combined the QUIJOTE MFI, WMAP, and \planck\ data. }
    \label{fig:mean_std_beta_s_data}
\end{figure}

Subsequently, we perform a T-T correlation analysis at \Nside\ = 32 between pairs of recovered Stokes maps corresponding to the two reconstructed synchrotron moments. 
To ensure the robustness of the correlation, we exclude some super-pixels from our analysis for which the Pearson correlation coefficient between the moment maps falls below 0.3. This prescription resulted in the removal of 0.5\% of the total pixels from the analysis. 
We compute the inverse-variance-weighted mean and the uncertainty, $\sigma_{\beta_s}^{\rm sys+stat}$, of the spectral indices, following Section~\ref{sec:TT_correlation}, from the estimated spectral index for 18 rotation angles ($\alpha \in [0^{\circ}, 85^{\circ}]$).
In the upper panel of Figure~\ref{fig:mean_std_beta_s_data}, we present the inverse-variance-weighted mean map of spectral index at \Nside=32. In the lower panel, we show the corresponding uncertainty map $\sigma_{\beta_s}^{\rm sys+stat}$.

\begin{figure}
    \centering
    \includegraphics[width=0.5\textwidth]{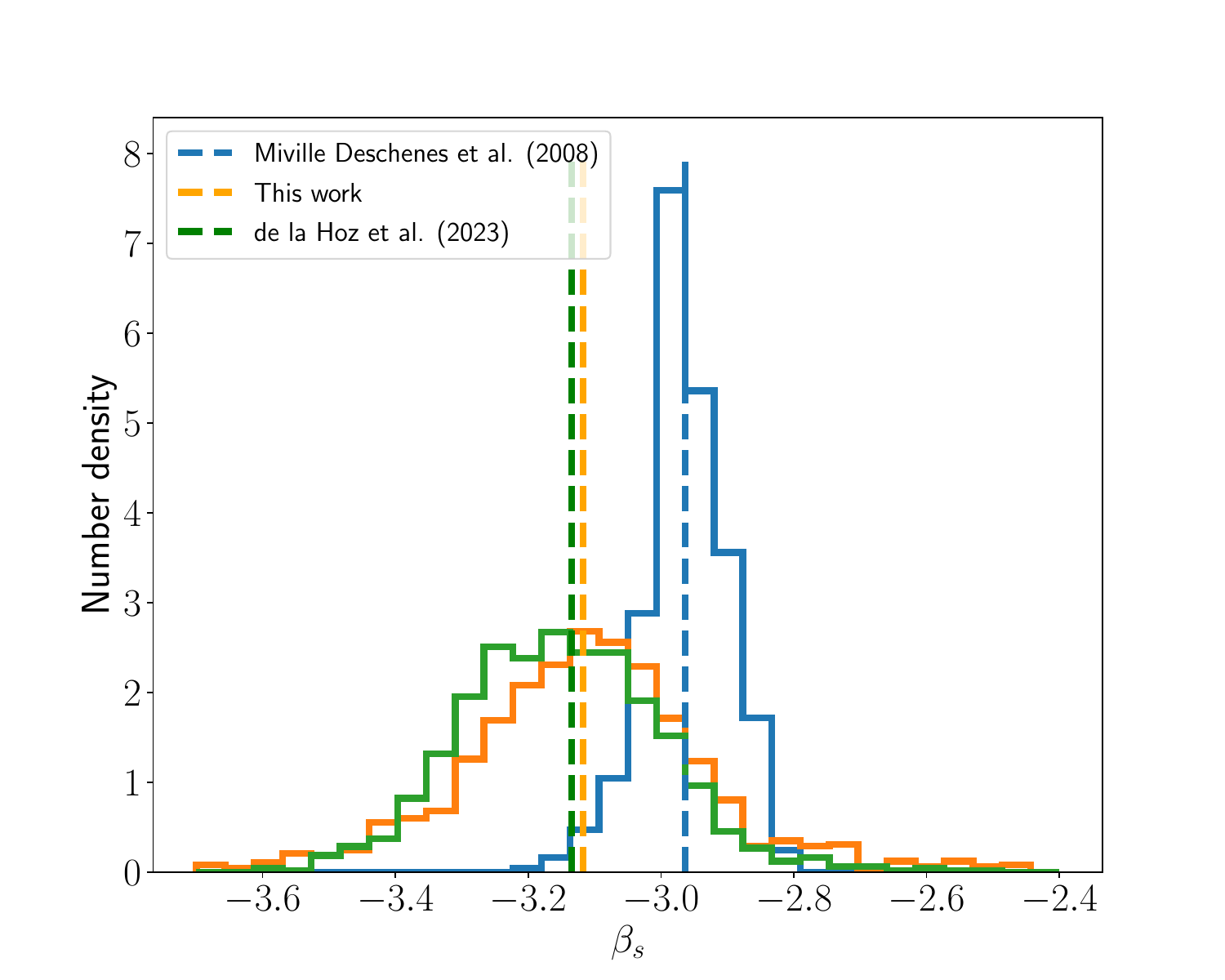}
    \caption{Distribution of the synchrotron spectral index from `Model 4' of \cite{Miville-Deschenes:2008} (blue), \cite{de_la_Hoz:2023} (green) and our estimation (orange). The vertical dashed lines correspond to the weighted mean values of the respective distributions. }
    \label{fig:hist_data_beta_s}
\end{figure}

The inverse-variance weighted mean and uncertainty of the  synchrotron spectral index within the Galactic plane region ($|b| < 10^{\circ}$) is found to be $\beta_s^{\text{plane}} = -3.05 \pm 0.01$, while at high Galactic latitudes which is dominated by the NPS for the choice of our mask, it steepens to $\beta_s^{\text{high-lat}} = -3.13 \pm 0.02$. This behaviour is consistent with earlier findings in the literature \citep{Fuskeland:2014, Krachmalnicoff:2018, Fuskeland:2021}, which suggest that the spectral index becomes steeper at high latitudes due to reduced depolarisation effects compared to those prevalent near the Galactic plane. It is important to note that the quoted uncertainties represent only the estimation errors arising from instrumental noise and systematics; they do not include the intrinsic dispersion of the spectral index, which we describe below. 
In Figure~\ref{fig:hist_data_beta_s}, we present a comparison of the spectral index distribution derived in this work (orange) with those reported by \cite{de_la_Hoz:2023} (green) and \cite{Miville-Deschenes:2008} (blue). The result in \cite{de_la_Hoz:2023} is based on fitting a parametric model to data from QUIJOTE MFI, WMAP, and \planck, including frequency channels up to 353 \GHz\ to improve the modelling of thermal dust emission. 
In contrast, our new method adopts a semi-blind strategy combined with a moment expansion approach, that makes fewer assumptions about SEDs of foregrounds and reduces reliance on detailed foreground parametrisation.
Furthermore, we exclude the higher-frequency channels ($\nu > 40$\GHz) to minimise contamination from dust emission and instead focus on the low-frequency regime ($\nu < 40$\GHz) where synchrotron emission dominates. Nevertheless, both independent approaches yield consistent results when applied to the data. The spectral index $\beta_s$ derived in \cite{Miville-Deschenes:2008} is 
widely adopted as a reference template for the synchrotron emission modelling and is incorporated into the standard \pysm\ models. This $\beta_s$ map was estimated using 408 \MHz\ Haslam map \citep{1982A&AS471H} and WMAP 23 \GHz\ data using a model based on a Galactic magnetic field (`Model 4'). 
However, both the spectral index distributions obtained in this work and by \citet{de_la_Hoz:2023} exhibit noticeably larger dispersion compared to that reported by \citet{Miville-Deschenes:2008}. To quantify this broadening, we compute the inverse-variance-weighted mean and standard deviation of the spectral index for all three $\beta_s$ maps over the analysed region. The estimated mean and standard deviation due to intrinsic scatter are $-2.96$ and $0.06$ for \citet{Miville-Deschenes:2008}, $-3.11$ and $0.21$ for this work (a broadening by a factor of $\sim$3.5), and $-3.13$ and $0.16$ for \citet{de_la_Hoz:2023} (a broadening by a factor of $\sim$2.7), respectively. A similarly broad variation in spectral index was also reported in \cite{Weiland:2022} over comparable sky regions. Additional studies \citep{Fuskeland:2014,10.1093/mnras/stv1328,  Krachmalnicoff:2018, Fuskeland:2021} further support the presence of significant spatial variability in the synchrotron spectral index across the sky.

\begin{figure*}
    \includegraphics[width=\textwidth]{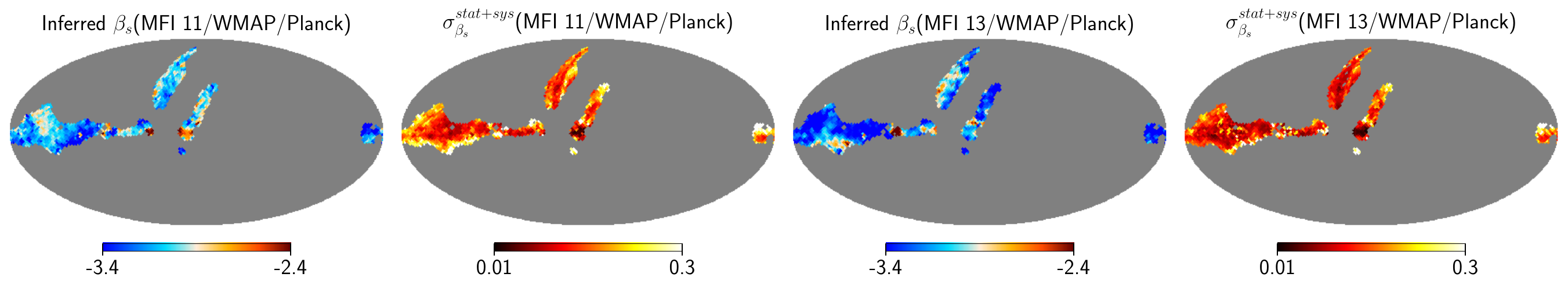}
    \caption{ Synchrotron spectral index and corresponding uncertainty maps obtained using the QUIJOTE MFI 11 \GHz\ map alone in replacement of QUIJOTE MFI combined maps of 11 and 13 \GHz\ are shown in the first and second panels, respectively. The similar plots of synchrotron spectral index and corresponding uncertainty maps when we use QUIJOTE MFI 13 \GHz\ data alone are shown in the third and fourth panels respectively. } 
    \label{fig:mean_std_beta_s_data_11_13}
\end{figure*}

We further conduct the analysis to evaluate the sensitivity of the derived synchrotron spectral index to the specific choice of QUIJOTE MFI input data. We examine how the results vary when the combined MFI map, considered as a baseline, is replaced with the individual frequency maps at 11 and 13 \GHz.  When we consider only 11 \GHz\ map in the component separation, the resulting spectral index has a mean and standard deviation of $-3.09$ and $0.23$ respectively, indicating a systematic shift of only 2.1\% ($\delta \beta_s = 0.07$) compared to the baseline result obtained with the combined MFI data. The corresponding spectral index and uncertainty maps are shown in the first and second  panels of Figure~\ref{fig:mean_std_beta_s_data_11_13}, respectively. A visual comparison with Figure~\ref{fig:mean_std_beta_s_data} reveals that the overall morphology remains largely consistent, indicating that the 11 \GHz\ map alone provides a stable recovery of the synchrotron spectral structure. In contrast, when the 13 \GHz\ map is considered in cPILC, the results show a significant deviation. The derived spectral index and uncertainty maps for this case are displayed in the third and fourth panels of Figure~\ref{fig:mean_std_beta_s_data_11_13}. These maps differ substantially from those obtained using either the combined MFI data or the 11 \GHz-only case, particularly in high S/N regions such as the Fan region and NPS. The mean and standard deviation of the spectral index in this scenario are found to be $-3.24 $ and $ 0.41$, respectively, with the increased dispersion reflecting the lower S/N of the 13 \GHz\ data. We revisited the simulations to verify the origin of the large intrinsic scatter in $\beta_s$ when including the 13~\GHz\ channel in the cPILC analysis. Our simulation results yield a mean and standard deviation of $-2.95$ and $0.27$, respectively, representing an increase in dispersion of approximately 68\% compared to the case using only the 11~\GHz\ channel. This finding suggests that the analysis based on the 13~\GHz\ channel alone may, in fact, degrade the reconstruction of the moment maps and consequently the reliability of the inferred spectral index due to its lower S/N. On the other hand, the analysis based on 11 \GHz\ alone or combined maps 11 and 13 \GHz\  give consistent results. Therefore, we adopt the joint use of both QUIJOTE MFI channels as default for analysis in the subsequent  section. 

\section{Analysis over regions}\label{sec:reg_analysis}
\begin{figure}
    \centering
    \includegraphics[width=0.5\textwidth]{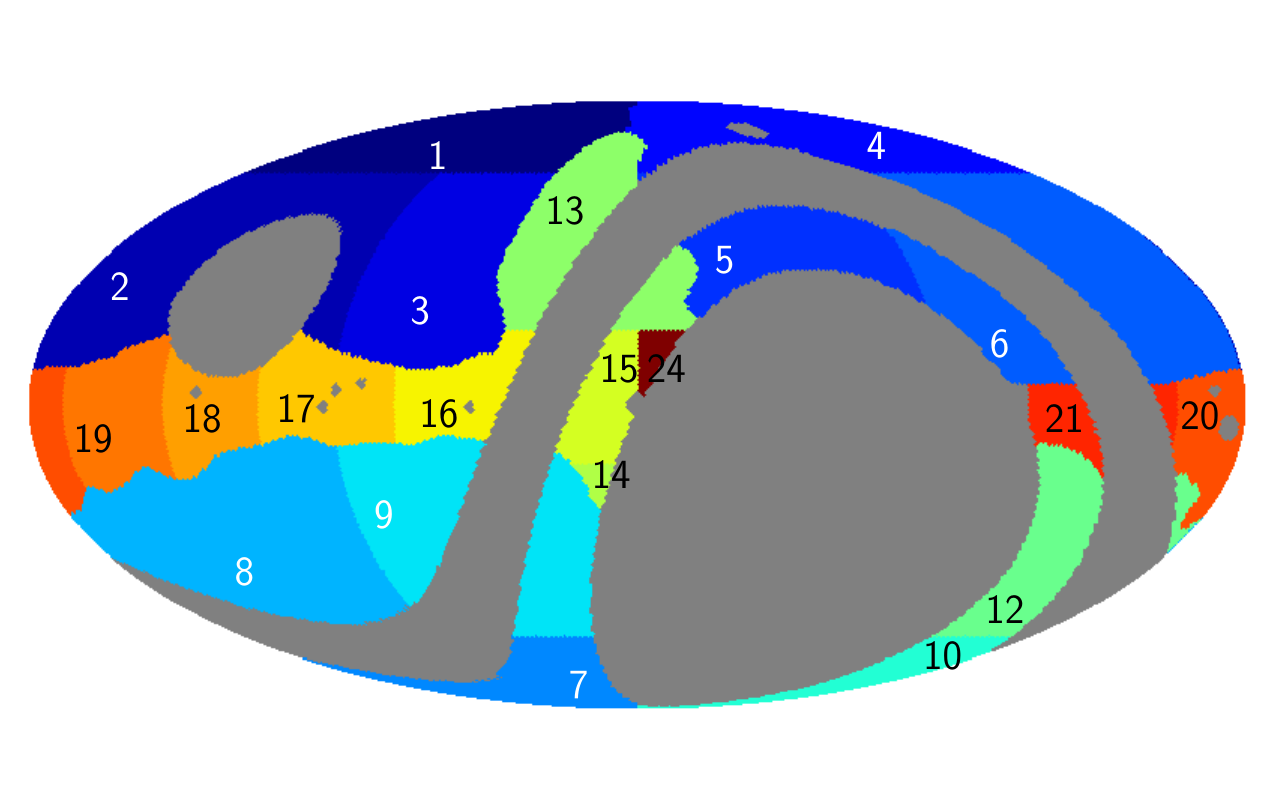}
    \caption{Regions used for estimation of the spectral index defined in \cite{Fuskeland:2014}. The grey regions are masked since the regions are not covered by QUIJOTE MFI survey.  Some of the bright sources are masked  using P06 polarisation mask provided by WMAP.} 
    \label{fig:region_map}
\end{figure}

\begin{figure}
    \includegraphics[width=\linewidth]{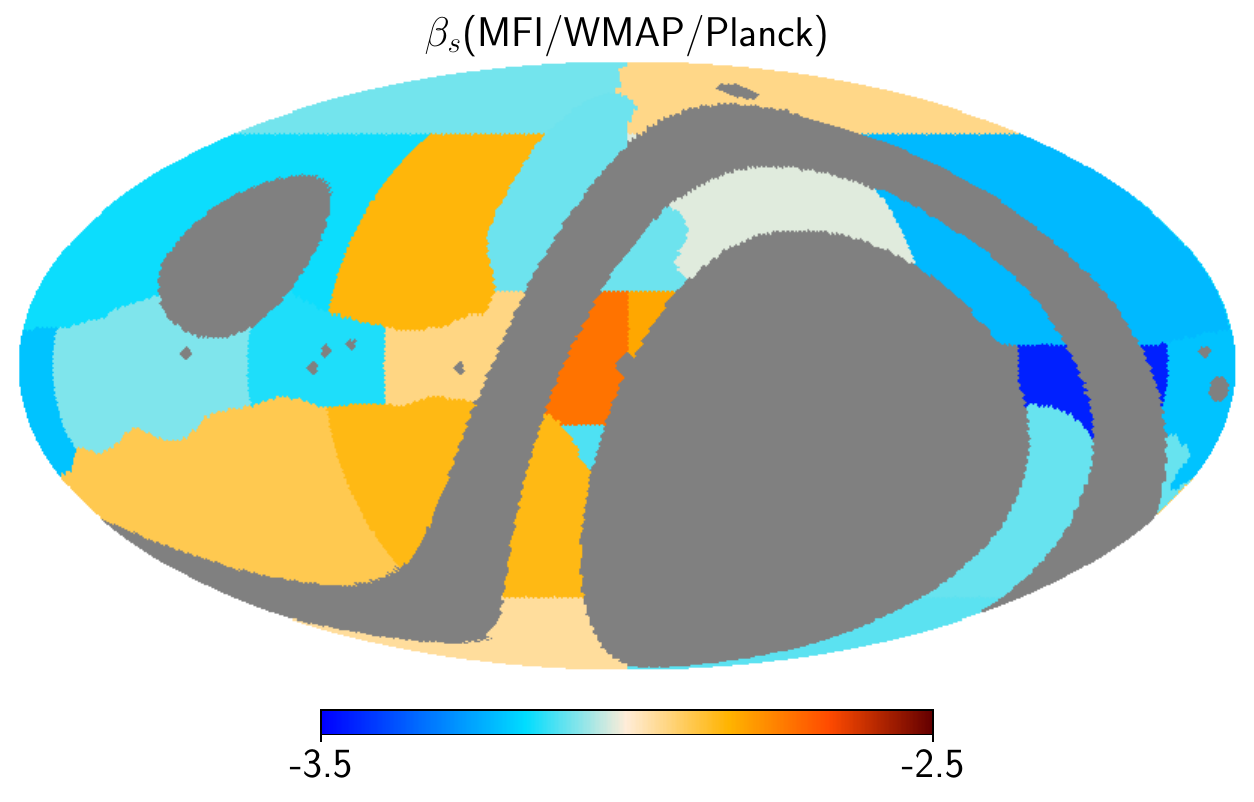}
    \includegraphics[width=\linewidth]{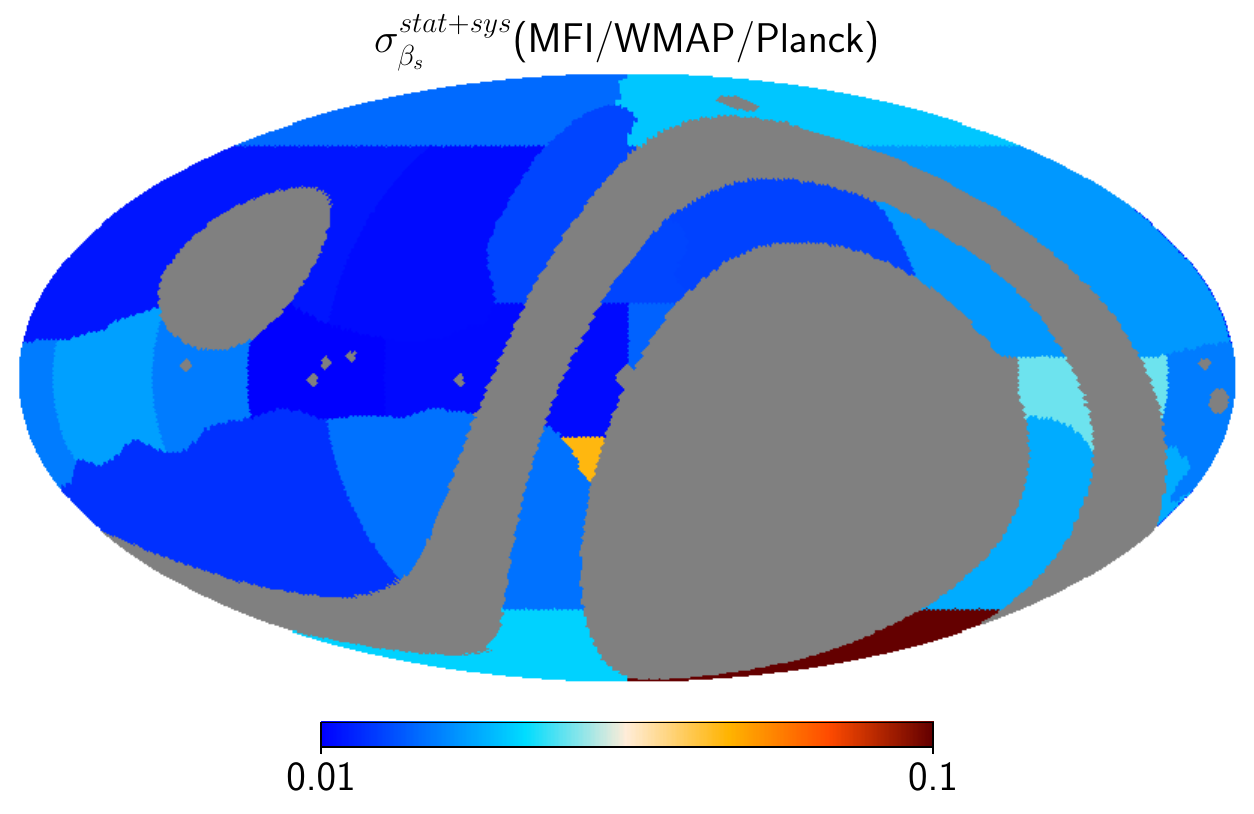}
  
   \caption{Upper panel: The derived spectral index over the regions. Lower panel: The uncertainty map of the derived spectral index.}
   \label{fig:beta_maps_regions_data}
\end{figure}

We extend the analysis to the low S/N regions of QUIJOTE MFI survey  partitioning the sky into regions defined in \cite{Fuskeland:2014}. We multiply the QUIJOTE MFI survey mask \citep{Rubiño-Martín:2023} by the mask used in \cite{Fuskeland:2014}, thereby redefining the effective sky coverage within each subregion. As a result, several regions, specifically regions 11, 22, and 23 as defined in Figure 1 of \cite{Fuskeland:2014} are excluded from our analysis. In addition, we apply further masking to exclude some of the brightest and most contaminated areas using P06 polarisation mask of WMAP discussed in Section~\ref{sec:data}. The final analysis mask  is shown in Figure~\ref{fig:region_map}. We perform the spectral index estimation in each of these regions following the same methodology used in the previous section. In this analysis, we use the combined map of QUIJOTE MFI 11 and 13 \GHz\ data in our component separation procedure. Subsequently, the T-T correlation between moments are performed over the sub-pixels of each region following same procedure as in Section~\ref{sec:TT_correlation}. The inverse-variance-weighted mean and uncertainty of the derived spectral index values for each region are reported in Table~\ref{table:1}, and the corresponding maps are presented in Figure~\ref{fig:beta_maps_regions_data}. It is important to note that the synchrotron spectral index varies significantly within these broad regions, as seen in our earlier results. 
Therefore, the average spectral indices obtained by fitting the synchrotron moments are of limited interpretative value when spatial variability is ignored. This is because the underlying assumption of T–T plotting is that all pixel pairs within a region share a common spectral index, which may not hold across all regions. In the literature, several studies have conducted T–T correlation analyses using two frequency data over similar regions with \planck\ and WMAP data \citep{Fuskeland:2014, Fuskeland:2021, Watts:2024}. Their results are also typically scattered due to such internal variations.

\begin{figure}
    \includegraphics[width=\linewidth]{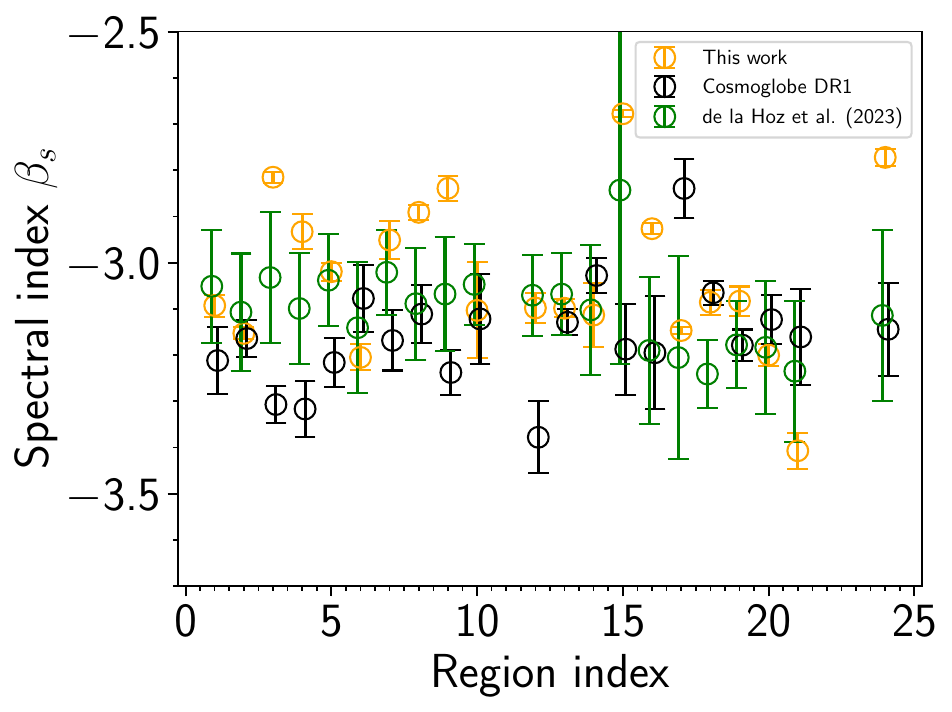}
   \caption{Synchrotron spectral index as a function of the region number obtained in this work (orange), \cosmoglobe\ \ensuremath{\tt DR1} (black), and \cite{de_la_Hoz:2023} (green).  The data shown in green are the inverse-variance-weighted mean of spectral index and uncertainties computed from $\beta_s$ and $\sigma_{\beta_s}$ maps of \cite{de_la_Hoz:2023} over regions.}
   \label{fig:beta_compare_regions_data}
\end{figure}

We compare our results with the recent findings of the \cosmoglobe\ \ensuremath{\tt DR1}\footnote{\url{https://www.cosmoglobe.uio.no/products/cosmoglobe-dr1.html}} \citep{Watts:2024} and \quijote\ \citep{de_la_Hoz:2023}. \cite{Watts:2024} performed T–T correlation analyses over comparable sky regions using \cosmoglobe\ maps at K-band and 30 \GHz\ \citep{cosmoglobe:2023}. \cite{de_la_Hoz:2023} performed the parametric fitting using QUIJOTE MFI 11, 13 \GHz,
WMAP 9-year K, Ka, \planck\ PR4 data. In Figure~\ref{fig:beta_compare_regions_data}, we display the spectral index obtained in our analysis (orange), and those found in  \cite{de_la_Hoz:2023} (green) and \cite{Watts:2024} (black). Since \cite{de_la_Hoz:2023} obtained the results from pixel-by-pixel fitting the data instead of joint fitting to all pixels in the regions of Figure~\ref{fig:region_map}, the green points represented in Figure~\ref{fig:beta_compare_regions_data}  are the inverse-variance weighted average of the spectral indices of \cite{de_la_Hoz:2023} at all pixels enclosed in each region. The corresponding plotted error bars are the weighted standard deviation estimated over the same set of pixels, and this explains why they are typically larger. The differences in results  from the three methods might be due not only to differences in the datasets used but to the different methodologies. For instance, the $\beta_s$ values in parametric fitting approach of \cite{de_la_Hoz:2023} is prior dominated especially in low S/N regions, which can produce noticeable discrepancies. Also, pixel-by-pixel fitting instead of fitting the data over entire region, could cause differences. However, we observe reasonable consistency, especially at low-Galactic latitudes. Next, we investigate the low-Galactic regions  which has a large overlapping region between our mask and the mask used in \cite{Watts:2024}.  We find that our estimated spectral indices are in very good agreement with those reported by \cite{Watts:2024} in regions 13 (NPS), 14 (Galactic centre), 18, 19 (Fan region), and 20. The rest of the Galactic regions (15, 16, 17, and 21) show some level of disagreement between the two analyses. In these low-Galactic regions, we find that the spectral indices obtained in \cite{Watts:2024} exhibit larger uncertainties than ours. The improvement in precision in our analysis arises from the inclusion of low-frequency data from the QUIJOTE MFI survey, which significantly enhances sensitivity to the synchrotron component. In particular, we find steeper spectral indices than \cite{Watts:2024} in regions 15, 16, and 24, while in region 17, our estimate yields a slightly flatter index.  At higher Galactic latitudes, however, the comparison becomes less straightforward. 
This is primarily due to the smaller overlap between the regions of patches used in our analysis and those in \cite{Watts:2024} at high Galactic latitudes, and second, due to the low S/N data in these regions. As a result, direct comparison of the results in those patches is less reliable.
To capture the overall trend, we compute the inverse-variance weighted mean and uncertainty of the spectral index across the Galactic plane (regions 15–25) and at high Galactic latitudes (regions 1–14), obtaining $\beta_{s}^{\mathrm{plane}} = -2.96 \pm 0.01$ (stat + sys) and $\beta_{s}^{\mathrm{high\text{-}lat}} = -3.02 \pm 0.02$  (stat + sys), respectively. The value within the Galactic plane is in good agreement with previous results from \cite{Fuskeland:2014} and \cite{Watts:2024}. At high Galactic latitudes, however, our estimate is slightly less steep than those reported in \cosmoglobe\ and \cite{Fuskeland:2014}. This mild difference may be attributed to the different sky coverage and inclusion of low-frequency QUIJOTE MFI data in our analysis, which improves synchrotron sensitivity.

\begin{table}[h!]
\centering
\caption{Synchrotron spectral index measured for each region. }
\begin{tabular}{p{4.2cm} p{2.8cm}} 
 \hline
 Region & Spectral index\\ 
 \hline\hline

1  & $-3.09 \pm 0.02$ \\
2  & $-3.15 \pm 0.01$ \\
3  & $-2.81 \pm 0.01$ \\
4  & $-2.93 \pm 0.03$ \\
5  & $-3.01 \pm 0.01$ \\
6  & $-3.20 \pm 0.02$ \\
7  & $-2.95 \pm 0.04$ \\
8  & $-2.89 \pm 0.01$ \\
9  & $-2.83 \pm 0.02$ \\
10 & $-3.10 \pm 0.10$ \\
12 & $-3.09 \pm 0.03$ \\
13 & $-3.09 \pm 0.01$ \\
14 & $-3.11 \pm 0.06$ \\
15 & $-2.67 \pm 0.01$ \\
16 & $-2.92 \pm 0.01$ \\
17 & $-3.14 \pm 0.01$ \\
18 & $-3.08 \pm 0.02$ \\
19 & $-3.08 \pm 0.03$ \\
20 & $-3.20 \pm 0.02$ \\
21 & $-3.40 \pm 0.03$ \\
24 & $-2.77 \pm 0.01$ \\
\hline
 Mean& $-2.99 \pm 0.02$\\
 \hline
\end{tabular}

\label{table:1}
\end{table}


\section{Summary and conclusion}\label{sec:summery and conclusion}
We have developed a new approach based on constrained-ILC and moment expansion of CMB foregrounds that allows us to recover synchrotron moment maps, enabling the construction of a synchrotron spectral index map through their correlation. This approach is fully developed on map-based semi-blind component separation and differs from parametric fitting of a model to the data. The analysis incorporates 2$^\circ$ smoothed data from QUIJOTE MFI at 11 and 13 \GHz, WMAP 9-year data at K and Ka bands, and the \planck\ PR3 data at LFI 30 GHz channel. These frequencies are dominated by synchrotron emission, making them particularly suitable for our study. We first validate the method using simulations and demonstrate that it can recover the synchrotron spectral index at high fidelity in the regions with high signal-to-noise ratio. The inferred results remain largely unbiased in the presence of polarised AME with a polarisation fraction below the current observational upper limit. Then we apply the method to real data to reconstruct the synchrotron spectral index map at the Galactic plane and the NPS region. Over this region our estimated spectral index map is consistent with the one obtained in \cite{de_la_Hoz:2023} which used a parametric model fitting to a similar data sets. Our analysis differs from that of \cite{de_la_Hoz:2023} in two key aspects. First, we restrict our dataset to channels below 40 \GHz, whereas \cite{de_la_Hoz:2023} employed all polarised channels up to 353 \GHz\ to constrain dust parameters. Second, we combine the QUIJOTE MFI  11 \GHz\ and 13 \GHz\ maps since the noise in these two MFI channels is significantly correlated, which can compromise the performance of our semi-blind component separation technique. Instead of the combined QUIJOTE MFI maps, using only the 11 \GHz\  data leaves the results remain approximately the same with only 2.1\% drift of mean $\beta_s$. However, when we use  QUIJOTE MFI 13 \GHz\ data, the obtained spectral index gets largely dispersed due to low S/N at 13 \GHz\ channel. We find the inverse-variance-weighted mean and dispersion of the spectral index distribution are $-3.11$ and $0.21$ respectively with combined QUIJOTE MFI data. This variability is notably larger than that predicted by the widely used synchrotron template from \cite{Miville-Deschenes:2008}, indicating that this template may underestimate spatial variation in the synchrotron spectral index due to many assumptions in synchrotron modelling and significant smoothing. We find moderately steeper spectral index at high Galactic latitude compared to the Galactic plane, 
consistent with previous studies in the literature.\\

We extend our analysis across the full QUIJOTE MFI survey region by dividing the sky into 21 subregions. This analysis yields a inverse-variance-weighted mean and uncertainties of the synchrotron spectral index over all regions of $\beta_s^{\rm all-region} = -2.99 \pm 0.01$. Considering the Galactic plane and high-latitude regions separately, we find the values are $\beta_s^{\rm plane} = -2.96 \pm 0.01 $ and  $\beta_s^{\rm high-lat} = -3.02 \pm 0.02 $ respectively. These results are consistent with previous estimates reported in \cite{Fuskeland:2014} and \cite{Watts:2024}, which performed T-T correlation analyses over similarly defined sky patches. We also compare our estimates on a patch-by-patch basis with the inverse-variance weighted average of the spectral indices from \cite{de_la_Hoz:2023}, and the spectral index values reported in \cite{Watts:2024}. Although the results from three methods are similar at some low-Galactic high S/N regions, there is a difference in high-Galactic regions due to lower sensitivity of the data and  differences in the analysis techniques. 

In this analysis, we assume that the synchrotron spectrum follows a simple power-law. If the true synchrotron spectrum exhibits a significant positive curvature, the second-order moment, as described in \cite{Remazeilles:2021MNRAS.503.2478R}, would become relevant. However, our current study does not account for such higher-order effects. Accurate estimation of higher-order moments will require data with improved sensitivity across a larger number of low-frequency channels. Future experiments such as QUIJOTE-MFI2 \citep{MFI2, mfi2:2024} and \textit{LiteBIRD} \citep{LiteBIRD:2023} will provide the necessary frequency coverage and sensitivity to better constrain the synchrotron curvature, thereby extending the applicability of this technique.

This work presents the first application of a semi-blind component separation technique for estimating maps of foreground spectral indices, offering an alternative to fully parametric model-fitting approaches. Our method mitigates bias from unwanted components by deprojecting their first few moments using partial knowledge of their SEDs, thereby enhancing robustness to model inaccuracies. 
The advantage of the method proposed here is the use of the moment expansion approach, which makes minimal assumptions on the SED of the foregrounds, as compared to the existing parametric methods.

\section*{Data availability}
All the synchrotron moment maps and the derived spectral index maps obtained in this work are publicly available in the  \faGithub\href{https://research.iac.es/proyecto/quijote/}{QUIJOTE web page}. The data in Table~\ref{table:1} is available in electronic form at the \faGithub\href{http://cdsweb.u-strasbg.fr/cgi-bin/qcat?J/A+A/}{CDS}.

\begin{acknowledgements} 
We thank the staff of the Teide Observatory for invaluable assistance in the commissioning and operation of QUIJOTE.
The QUIJOTE experiment is being developed by the Instituto de Astrofisica de Canarias (IAC), the Instituto de Fisica de Cantabria (IFCA), and the Universities of Cantabria, Manchester and Cambridge. 
Partial financial support was provided by the Spanish Ministry of Science and Innovation
under the projects AYA2007-68058-C03- 01, AYA2007-68058-C03-02, 
AYA2010-21766-C03-01, AYA2010-21766-C03- 02, AYA2014-60438-P, 
ESP2015-70646-C2-1-R, AYA2017-84185-P, ESP2017- 83921-C2-1-R, 
PGC2018-101814-B-I00, PID2019-110610RB-C21, PID2020-120514GB-I00, 
IACA13-3E-2336, IACA15-BE-3707, EQC2018-004918-P, 
PID2023-150398NB-I00, PID2023-151567NB-I00,  PID2022-139223OB-C21 and PID2022-140670NA-I00, 
the Severo Ochoa Programs SEV-2015-0548 and CEX2019-000920-S, 
the Maria de Maeztu Program MDM-2017-0765, and by the Consolider-Ingenio project CSD2010-00064 (EPI: Exploring the Physics of Inflation). 
We acknowledge support from the ACIISI, Consejeria de Economia, Conocimiento y Empleo del Gobierno de Canarias and the European Regional Development Fund (ERDF) under grant with reference ProID2020010108, and Red de Investigaci\'on RED2022-134715-T funded by MCIN/AEI/10.13039/501100011033. 
This project has received funding from the European Union's Horizon 2020 research and innovation program under grant agreement number 687312 (RADIOFOREGROUNDS), and the Horizon Europe research and innovation program under GA 101135036 (RadioForegroundsPlus). MFT acknowledges support from the Enigmass+ research federation (CNRS, Université Grenoble Alpes, Université Savoie Mont-Blanc). FP acknowledges support from the MICINN under grant numbers PID2022-141915NB-C21. EMG acknowledges support from the Plan
Complementario AstroHEP funded by the "European Union
Next GenerationEU/PRTR" and the Government of the
Autonomous Community of Cantabria.
Some of the presented results are based on observations obtained with \planck\ (\url{http://www.esa.int/Planck}), an ESA science mission with instruments and contributions directly funded by ESA Member States, NASA, and Canada. We acknowledge the use of the Legacy Archive for Microwave Background Data Analysis (LAMBDA). Support for LAMBDA is provided by the NASA Office of Space Science. Some of the results in this paper have been derived using the Healpy and {\tt HEALPix} packages \citep{Healpix, Healpix2}. The paper uses the python packages, Numpy \citep{Numpy:2020}, Matplotlib \citep{Matplotlib:2007}, Scipy \citep{scipy:2020}. DA thanks \cosmoglobe\ project (\url{https://www.cosmoglobe.uio.no/products}) for making their spectral index map and masks publicly available. 
\end{acknowledgements} 

\bibliographystyle{aa}
\bibliography{quijote_sync}

\clearpage

\appendix  
\onecolumn
\section{Analysis using PR4 data at 30 GHz}\label{sec:pr4_results}
In this section, we present a revised analysis incorporating the Planck PR4 data at 30 GHz, while all other frequency data remain the same as described in the main text. Figure~\ref{fig:beta_s_mean_std_pr4} shows the recovered moment maps. Compared to the results presented in Figure~\ref{fig:moment_maps_data}, the moment maps obtained using the PR4 data show no major differences relative to those derived using the PR3 data in the main analysis. Using the TT correlation between the recovered moment maps, we derive the $\beta_s$ spectral index map and its corresponding standard deviation map, also shown in Figure~\ref{fig:beta_s_mean_std_pr4}. The  inverse-variance-weighted  mean and dispersion of the derived spectral index map are --3.12 and 0.22, respectively. These values indicate no significant change relative to the results reported in the main text.
\begin{figure*}[htbp]
\centering
\includegraphics[width=\textwidth]{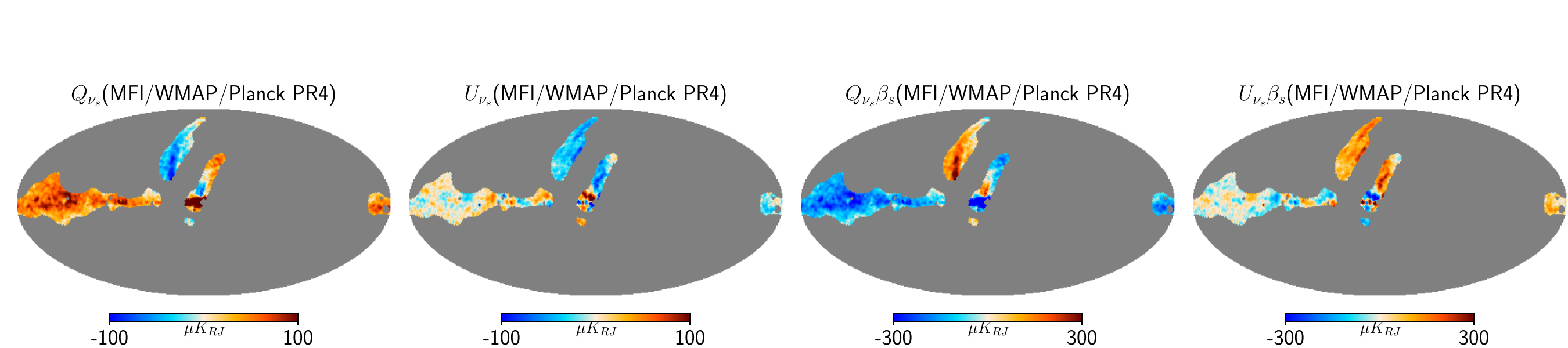}

      \caption{Synchrotron Stokes $Q_{\nu_s}$ (first panel), and $U_{\nu_s}$ (second panel) maps at 22.8 \GHz\ recovered from the cPILC component separation method using the QUIJOTE MFI 11 and 13 \GHz\ combined map, WMAP K and Ka bands and \planck\  PR4 30\GHz. The Stokes parameters corresponding to the spectral index modulated synchrotron amplitudes, $Q_{\nu_s}\beta_s,U_{\nu_s}\beta_s$ obtained from component separation are shown in the third and fourth panels respectively. All maps are at FWHM=2$^\circ$ at \Nside = 64.  }
    \label{fig:moment_maps_data_pr4}
\end{figure*}
\begin{figure*}[!htbp]
    \includegraphics[width=0.5\linewidth]{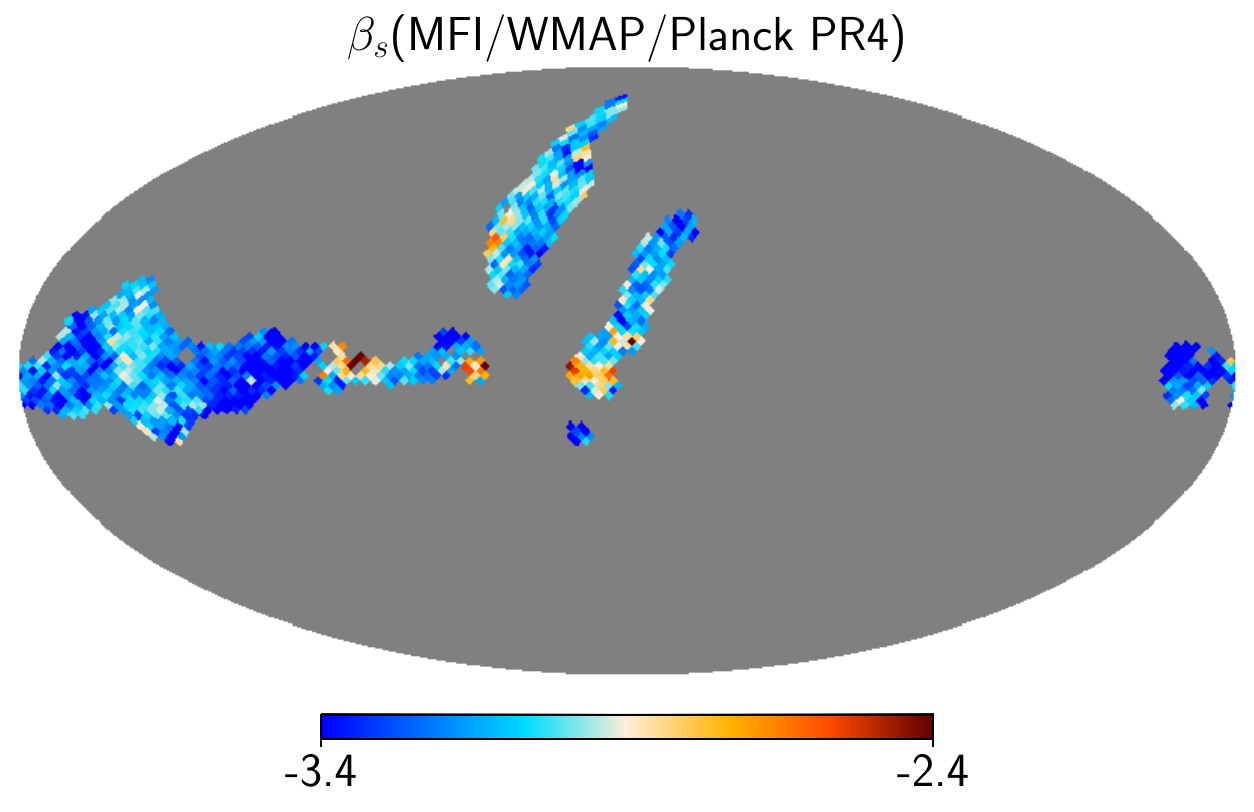}
    \includegraphics[width=0.5\linewidth]{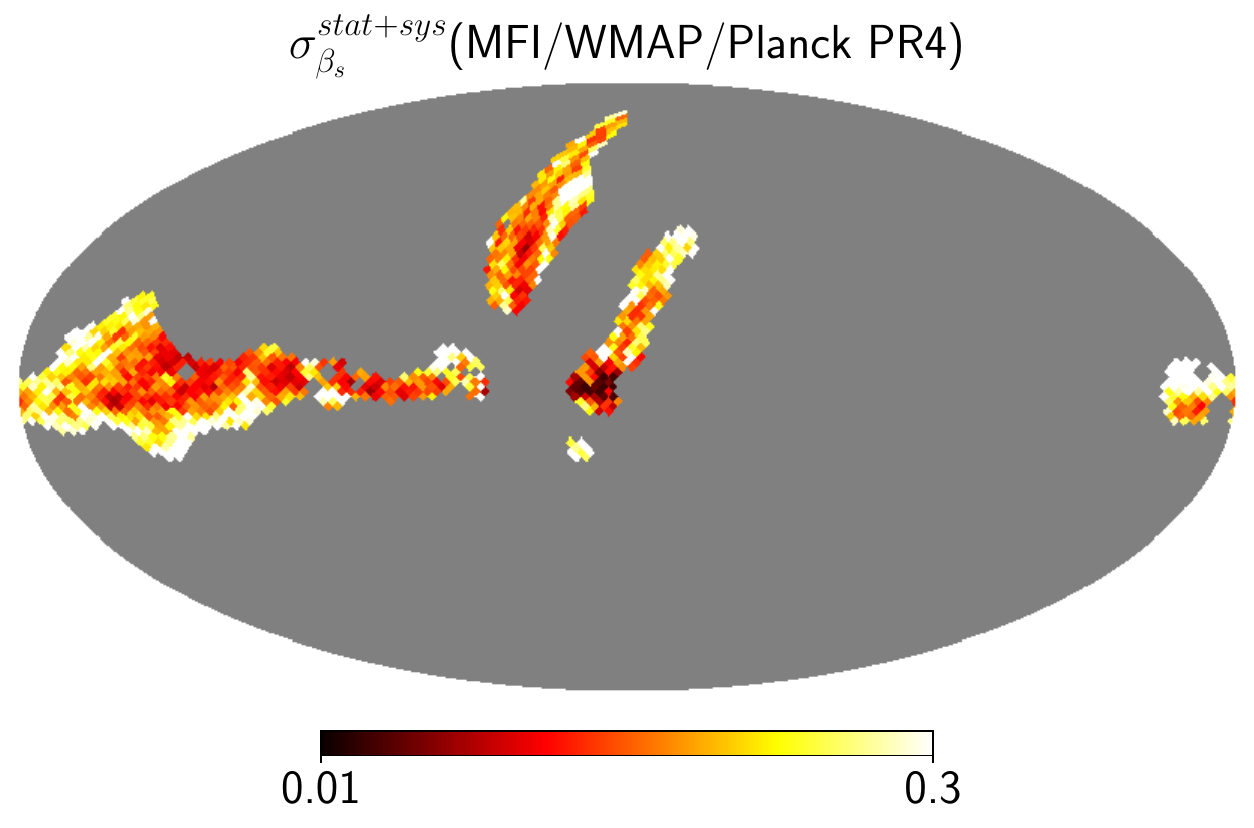}
    \caption{Synchrotron spectral index (left panel) and uncertainty (right panel) maps  obtained from the combined QUIJOTE MFI, WMAP, and \planck\ PR4 data.}
    \label{fig:beta_s_mean_std_pr4}
\end{figure*}


\end{document}